\documentclass[useAMS,usenatbib]{mn2e} 
\usepackage{aurelie} 
\usepackage{array} 
\usepackage{graphicx}
 \usepackage{natbib} 
\usepackage{pjournal} 
\usepackage{rotating} 
\usepackage{multirow}
\usepackage{amssymb}
\title [NG of the CIB anisotropies]{Non-Gaussianity of the Cosmic Infrared Background anisotropies II~: Predictions of the bispectrum and constraints forecast} 
\author[A. P\'enin, F. Lacasa and N. Aghanim]
{A. P\'enin$^{1}$ \thanks{E-mail: aurelie.penin@oamp.fr (AVR)},
F. Lacasa$^{2}$
and N. Aghanim$^{2}$\\
 $^{1}$ Aix Marseille Universit\'e, CNRS, LAM (Laboratoire d'Astrophysique de Marseille) UMR 7326, 13388, Marseille, France\\
  $^{2}$ Institut d'Astrophysique Spatiale (IAS), B\^atiment 121, F-91405 Orsay
(France); Universit\'e Paris-Sud 11 and CNRS (UMR 8617)}

\begin{document} \date{Accepted 0000. Received 0000}

\pagerange{\pageref{firstpage}--\pageref{lastpage}} \pubyear{2013}

\maketitle

\label{firstpage}


\begin{abstract}

Using a full analytical computation of the bispectrum based on the halo model together with the halo occupation number, we derive the bispectrum of the cosmic infrared background (CIB) anisotropies that trace the clustering of dusty-star-forming galaxies. We focus our analysis on wavelengths in the far-infrared and the sub-millimeter typical of the {\tt Planck/HFI} and {\tt Herschel/SPIRE} instruments, 350, 550, 850, and 1380 \um. We explore the bispectrum behaviour as a function of several models of evolution of galaxies and show that it is strongly sensitive to that ingredient. Contrary to the power spectrum, the bispectrum, at the four wavelengths, seems dominated by low redshift galaxies. Such a contribution can be hardly limited by applying low flux cuts. We also discuss the contributions of halo mass as a function of the redshift and the wavelength, recovering that each term is sensitive to a different mass range. Furthermore, we show that the CIB bispectrum is a strong contaminant of the Cosmic Microwave Background bispectrum at 850 \um~and higher.  Finally, a Fisher analysis of the power spectrum, bispectrum alone and of the combination of both shows that degeneracies on the HOD parameters are broken by including the bispectrum information, leading to tight constraints even when including foreground residuals.

\end{abstract}

\begin{keywords} Infrared : galaxies - Cosmology : large scale structure of the Universe - galaxies: statistics – \end{keywords}


\section{Introduction} Observations in the far-Infrared (FIR) and in the sub-millimeter range are limited by confusion effects. Extragalactic sources below the detection limit lead to brightness fluctuations, because of the low resolution of the instruments. The resulting Cosmic Infrared Background (CIB) contains the radiation emitted by dusty star-forming galaxies (DSFG) since the decoupling. These galaxies have very high star formation rates (SFR) and are thus the main locations of star formation in the Universe. Most of their energy ($\sim95 \%$) is emitted in the IR as dust reprocesses and re-emits UV starlight in the IR.  Only 15 \% of the CIB is directly resolved into sources by {\tt Herschel/SPIRE} at 250 \um~\citep{2010AA...518L..21O} and this fraction decreases with increasing wavelength.  Statistical methods that permit to reach fluxes much lower than the confusion limit, for instance stacking, enable to increase the fraction of resolved CIB to 73 \% at 250~\um~with {\tt Herschel/SPIRE} \citep{2012A&A...542A..58B}. The unresolved sources thus give access to the majority of the DSFG emission. The CIB displays anisotropies that are the consequence of the underlying spatial distribution of DSFG. These anisotropies further probe the clustering of DSFG and the link between galaxies and dark matter halos in the large scale structure. The redshift distribution of the sources constituting the CIB shifts towards higher redshifts as the wavelength increases \citep{2008A&A...481..885F}. As a result, a multi-wavelength study of the CIB gives, in principle, access to the redshift evolution of the anisotropies, and therefore to the evolution of the underlying population of galaxies.\\  The CIB anisotropies have been measured in the last few years over a wide range of scales and wavelengths from 100 \um~to 1380 \um. They first have been detected at 160~\um~\citep{2007ApJ...665L..89L} and in {\tt BLAST} data from 250 to 500 \um \citep{2009ApJ...707.1766V}. Several measurements have followed in the submillimeter range \citep{2010ApJ...718..632H,2010AAS...21538407F}. More recently, they have been measured in {\tt Herschel/SPIRE} data at 250, 350, and 500~\um~ \citep{2011Natur.470..510A,2012arXiv1208.5049V} and in {\tt Planck} data from 350 to 1380 \um~\citep{2011A&A...536A..18P} with unprecedented accuracy.  At lower wavelengths, the main limitation of a CIB measurement is the contamination by Galactic cirrus on large angular scales \citep{2011A&A...536A..18P}. \citet{2012A&A...543A.123P} measured the CIB power spectra at 100 and 160 \um~by accurately removing this component using an independent tracer of dust, namely neutral hydrogen. It enabled the authors to extend the range of scales of the CIB measurement.  \\
In parallel to the measurement of CIB, theoretical modeling has been the subject of a lot of activities. A biased linear power spectrum was fit to the data \citep{2007ApJ...665L..89L,2010ApJ...718..632H}, but a such a simple model has been ruled out by \citet{2011A&A...536A..18P}. More complex models of clustering dedicated to DSFG were developed and used to analyse the data. They combine two different models: one for the evolution of galaxies and one for the distribution of dark matter. Concerning the former, several recent models, parametric or physically motivated, that reproduce well current measurements of differential number counts and luminosity functions have been used \citep{2004ApJS..154..112L,2007MNRAS.377.1557N,2011A&A...529A...4B}.\\
For the distribution of galaxies versus dark matter, the halo model \citep{2002PhR...372....1C} has been applied to DSFG. It describes the spatial distribution of dark matter halos. The introduction of the halo occupation distribution (HOD) allows to extend this framework to the distribution of galaxies by prescribing the number of galaxies within a halo as a function of the halo mass. This analytical approach has been widely applied and used to study the CIB power spectrum \citep{2009ApJ...707.1766V,2012A&A...537A.137P,2012MNRAS.422.1324X}. \\ Being a fully analytical approach, the halo model can be extended to higher order correlations.  The 3-point correlation function as well as its Fourier space analogue, the bispectrum, is the lowest order indicator of non-Gaussianity. First attempts at predicting the bispectrum in the millimetre domain have been made by \citet{2003ApJ...598...86A} considering the measured two-point correlation function of both DSFG and radio galaxies together. The first measurement of the bispectrum of extragalactic point sources in {\tt WMAP1} data was performed by \citet{2003ApJS..148..119K}. At that time the correlated anisotropies of the CIB had not been detected, only the Poisson component had \citep{2000A&A...355...17L,2007ApJ...665L..89L}.  Moreover, at the frequencies probed by WMAP \citep{2003ApJ...583....1B} the dominant extragalactic sources are radio-emitting galaxies. They are unclustered which makes the bispectrum a constant. The subject of bispectrum from extragalactic sources has recently gained interest both from the point of view of the modeling, in particular for the CIB contribution \citep{2012MNRAS.421.1982L}, and from the point of view of the measurement \citep{2013arXiv1303.3535C}. This regain of interest has mostly two motivations. First, the CIB bispectrum is a complementary information to the number counts, luminosity function and power spectrum that could help constraining models of clustering and evolution of DSFG galaxies. Second, the CIB is a foreground to the CMB measurement and its non-Gaussianity needs to be looked at carefully.  \\
In this paper, we use the formalism developed in a companion paper \citet{Lacasaetpenin_bl}, referred to hereafter as Paper1, to compute predictions of the bispectrum of CIB anisotropies at far-infrared and millimeter wavelengths and to investigate the dependencies of the bispectrum, as a function of the wavelength. We summarize in Sect. \ref{par:formalism} the formalism of the bispectrum, detailed in Paper1. We summarize how we derive the predictions and investigate the dependencies of the CIB bispectrum on the input models of galaxies in Sect. \ref{par:pred+gal+hod}.  We then investigate, in Sect. \ref{par:zm_distrib}, the mass and redshift ranges of the dark matter halos in which DSFG are embedded and which contribute to the CIB bispectrum. We present the effect of the emissivity flux cuts in Sect. \ref{par:importance_fc}. Eventually, we compare the CIB bispectrum to that of the CMB in Sect. \ref{par:delta_fnl} and we carry a Fisher analysis to investigate to which extent combined data of power spectra and bispectra constrain the HOD parameters. Finally we conclude in Sect.~\ref{par:ccl}. 


\section{Formalism of the bispectrum}\label{par:formalism} 
We summarize here the formalism of the power spectrum and the bispectrum of CIB anisotropies. Both require two main ingredients, the emissivities that rule the evolution of DSFG and the distribution of dark matter halos in which galaxies are embedded, described by the halo model. We will introduce the halo occupation number, that rules how galaxies populate halos.
 \subsection{The angular power spectrum} 
Following \citet{2000ApJ...530..124H}, \citet{2001ApJ...550....7K}, \citet{2012A&A...537A.137P}, and using the Limber approximation, the angular power spectrum of the CIB anisotropies at a given wavelength $\lambda$ is:
\be
 C_\ell^\lambda = \int\frac{\dd z}{r^2}\frac{\dd r}{\dd z} a^2(z) \overline{j}^2_\lambda(z)P_{\mathrm{gal}}(k,z) + C_\ell^\mathrm{shot}
\label{eq:cl}
\ee
where $r$ is the comoving distance and $a(z)=(1+z)^{-1}$ is the scale factor. $\overline{j}_\lambda(z)$ is the mean emissivity of DSFG given by a model of evolution of galaxies (see Sect \ref{sect:model_evolution}). $P_{\mathrm{gal}}(k,z)$ is the galaxy 3D power spectrum derived from the halo model. $C_\ell^\mathrm{shot}$ is the shot-noise contribution described in Sect.\ref{Sect:shot-noise}.

\subsection{The angular bispectrum}
 As derived in Paper1, in the Limber approximation the bispectrum of the CIB anisotropies at a given wavelength $\lambda$ is 
\be \label{Eq:bisp_3Dtoang}
\bleq^\lambda = \int \frac{\dd z}{r^4} \,\frac{\dd r}{\dd z}\, a^3(z) \,\overline{j}^3_\lambda(z)\, B_\mathrm{gal}^\mathrm{clus}(k_{123},z) 
+\bleq^\mathrm{shot}
\ee 
where $B_\mathrm{gal}(k_{123},z)$ is the galaxy 3D bispectrum derived in the framework of the halo model, and $\bleq^\mathrm{shot}$ is described in Sect.\ref{Sect:shot-noise}. For clarity, we write $(k_1,k_2,k_3)$ as $k_{123}$ in the following.

\subsection{The 3D bispectrum from the halo model} 
Similarly to the 3D power spectrum, the 3D bispectrum has both clustering and the shot-noise contributions (see Paper1 for details): 
\be 
B_\mathrm{gal}(k_{123},z) = B_\mathrm{gal}^\mathrm{clus}(k_{123},z) + B_\mathrm{gal}^\mathrm{shot}(k_{123},z) 
\ee

\subsubsection{The clustering terms}
 In the framework of the halo model, the bispectrum can be written as the sum of three components, 1-halo, 2-halo and 3-halo terms: \be B_\mathrm{gal}^\mathrm{clus}(k_{123},z) = B^\mathrm{1h}_\mathrm{gal}(k_{123},z)+ B^\mathrm{2h}_\mathrm{gal}(k_{123},z)+B^\mathrm{3h}_\mathrm{gal}(k_{123},z) \ee The 1-halo term is the contribution of three galaxies within the same halo and it dominates at small spatial scales. The 2-halo term is the case of two galaxies within one halo and a third galaxy in another halo. It mainly contributes at intermediate spatial scales. Finally, the 3-halo term is the contribution of three galaxies in three distinct halos and dominates at large spatial scales. \\ The 1-halo term of the bispectrum reads : \bea B^\mathrm{1h}_\mathrm{gal}(k_{123},z) = \int \dd M \, \frac{\langle N_\mathrm{gal} (N_\mathrm{gal}-1)(N_\mathrm{gal}-2)\rangle} {\overline{n}_\mathrm{gal}^3} \frac{\dd N_h}{\dd M} \,\\ \times |u(k_1 | M) \, u(k_2 | M) \, u(k_3 | M)| \eea where $\dd N_h/dM$ is the halo mass function, $u(k| M)$ is the normalized Fourier transform of the halo density profile, and $\langle N_\mathrm{gal}\rangle$ is the probability of having $N_\mathrm{gal}$ galaxies in a halo of mass $M$. The latter is described by the halo occupation distribution \citep{2004ApJ...609...35K}.  Finally, $\overline{n}_\mathrm{gal}$ is the mean number of galaxies given by: 
\be 
\overline{n}_\mathrm{gal}= \int \dd M \frac{\dd N_h}{\dd M}\,\langle N_\mathrm{gal} \rangle 
\ee 
In the following, we use the mass function of \citet{1999MNRAS.308..119S}, the \citet{1996ApJ...462..563N1996ApJ...462..563N} halo density profile, and the HOD is described in more details in Sect \ref{sect:hod}.

In order to simplify the equations for the 2- and 3-halo terms, we introduce the following notations
\be \mathcal{F}_1(k,z) = \int \dd M \frac{\langle N_\mathrm{gal}(M)\rangle}{\overline{n}_\mathrm{gal}(z)} \,\frac{\dd N_h}{\dd M} \, b_1(M) \, |u(k|M)| \ee \be \mathcal{F}_2(k,z) = \int \dd M \frac{\langle N_\mathrm{gal}(M)\rangle}{\overline{n}_\mathrm{gal}(z)} \,\frac{\dd N_h}{\dd M} \,b_2(M) \, |u(k|M)|
\ee
\bea \mathcal{G}_1(k_1,k_2,z) = \int \dd M \frac{\langle N_\mathrm{gal}(N_\mathrm{gal}-1)\rangle}{\overline{n}_\mathrm{gal}(z)^2} \,\frac{\dd N_h}{\dd M}\, b_1(M) \, \\
\times|u(k_1|M) \, u(k_2|M)|\nonumber
\eea
where $b_1(M)$ and $b_2(M)$ are respectively the first and second order halo biases associated to the chosen mass function.\\ 
The 2-halo term becomes:
\bea\label{eq:Bk2h} \nonumber B^\mathrm{2h}_\mathrm{gal}(k_{123},z) &=& \mathcal{G}_1(k_1,k_2,z) \; P_\mathrm{lin}(k_3 , z) \, \mathcal{F}_1(k_3,z) \\ \nonumber&& +\ \mathcal{G}_1(k_1,k_3,z)\; P_\mathrm{lin}(k_2 , z) \,\mathcal{F}_1(k_2,z) \\
 && +\ \mathcal{G}_1(k_2,k_3,z) \; P_\mathrm{lin}(k_1 ,z ) \, \mathcal{F}_1(k_1,z)
 \eea
 and the 3-halo term writes:
\bea\label{Eq:Bk3hwFnot} &&B^\mathrm{3h}_\mathrm{gal}(k_{123},z) =\mathcal{F}_1(k_1,z)\, \mathcal{F}_1(k_2,z)\, \mathcal{F}_1(k_3,z)\\
\nonumber &&\times \left[F_{12}^s(k_1,k_2) P_\mathrm{lin}(k_1,z) P_\mathrm{lin}(k_2,z) +\mathrm{perm.}\right] \\
\nonumber && +\ \mathcal{F}_1(k_1,z)\, \mathcal{F}_1(k_2,z)\, \mathcal{F}_2(k_3,z) \times P_\mathrm{lin}(k_1 , z)\, P_\mathrm{lin}(k_2 , z)\\
\nonumber && +\ \mathcal{F}_1(k_1,z)\, \mathcal{F}_2(k_2,z)\, \mathcal{F}_1(k_3,z) \times P_\mathrm{lin}(k_1 , z)\, P_\mathrm{lin}(k_3 , z)\\
&& +\ \mathcal{F}_2(k_1,z)\, \mathcal{F}_1(k_2,z)\, \mathcal{F}_1(k_3,z) \times P_\mathrm{lin}(k_2 , z)\, P_\mathrm{lin}(k_3 , z)\nonumber 
\eea
where $P_\mathrm{lin}(k,z)$ is the linear power spectrum, and $\mathrm{F}_{\alpha\beta}^s$ is the kernel due to the non-linearity of gravity at second-order in perturbation theory \citep{1984ApJ...277L...5F,2012JCAP...02..047G}:
\be
F_{\alpha \beta}^s = \frac{2 k_\gamma^4 -3(k_\alpha^4+k_\beta^4)+3k_\gamma^2(k_\alpha^2+k_\beta^2) +10 k_\alpha^2k_\beta^2}{28 k_\alpha^2k_\beta^2}
\ee
with $\gamma$ the third index.

\subsubsection{The shot noise terms}\label{Sect:shot-noise}
 In the case of the power spectrum, there is only one shot noise term that depends on the differential number counts and the flux cut $S_{\mathrm{cut}}$. Thus, it is independent of the HOD: 
\be 
C_{\ell}^{\mathrm{shot}} = \int_0^{S_{\mathrm{cut}}} S^2\frac{\dd N}{\dd S}\dd S 
\ee 
The shot noise term of the bispectrum has two components. 
\be 
\bleq^{\mathrm{shot}} = \bleq^{\mathrm{shot1g}} + \bleq^{\mathrm{shot2g}} 
\ee 
The 1-galaxy term is the correlation of one galaxy with itself three times and the 2-galaxy term is the correlation of one galaxy with itself and with another one (within the same halo or not).  Similarly to the power spectrum, the 1-galaxy term is written: 
\bea 
\bleq^{\mathrm{shot1g}}&=&\int_0^{S_\mathrm{cut}} S^3\frac{dN}{dS}dS\label{eq:shot1g}\\ 
&=& \int \dd z \frac{\dd r}{\dd z} a^3(z) \, j^{(3)}(z)
 \eea 
where $j^{(n)}$ is the n-th order emissivity: 
\be
 j^{(n)}(z) = \frac{(1+z)^n}{\frac{\dd r}{\dd z}} \int_0^{S_\mathrm{cut}} S^n \frac{\dd^2 N}{\dd S \,\dd z} \dd S 
\ee 
The 2-galaxy shot noise term is:
 \bea 
\bleq^\mathrm{shot2g}& =& \int \frac{\dd z}{r^4} \frac{\dd r}{\dd z} a^3(z) \, j^{(1)}(z) \, j^{(2)}(z) \\ 
&\times& \left[ P_{\mathrm{gal}}(k_1)+P_{\mathrm{gal}}(k_2)+P_{\mathrm{gal}}(k_3) \right] 
\eea 
where $P_{\mathrm{gal}}(k)$ is the 3D power spectrum derived from the halo model. Thus this term depends on the HOD.

\subsection{The halo occupation distribution}\label{sect:hod} 
The standard halo occupation distribution rules the number of galaxies in a halo as a function of the halo mass only \citep{2004ApJ...609...35K,2010ApJ...719...88T}. In such a framework, the distribution of galaxies within a halo is Poissonian.  Simulations, as well as recent data suggest a necessary distinction between the major galaxy that lies in the centre of the halo and the satellite galaxies that populate the rest of the halo \citep{2002ApJ...575..587B,2004ApJ...609...35K,2005ApJ...633..791Z,2007MNRAS.376..841V}. Thus \be \langle N_{\mathrm{gal}}\rangle =\langle N_{\mathrm{cen}}\rangle+\langle N_{\mathrm{sat}}\rangle\ . \ee According to the prescription of \cite{2010ApJ...719...88T}, the occupation function of central galaxies is 
\be 
\langle N_{\mathrm{cen}}\rangle = \frac{1}{2}\left[1+\mbox{erf}\left(\frac{\log M-\log M_{\mathrm{min}}}{\sigma_{\log M}}\right)\right]
\label{eq:hod_cent} 
\ee 
where $M_\mathrm{min}$ is the halo mass at which a halo has a 50\% probability of hosting a central galaxy and $\sigma_{\log M}$ controls the width of the transition between zero and one central galaxy.  There is a smooth transition between low-mass halos that do not contain bright enough galaxies ($M<<M_\mathrm{min}$) and more massive halos that always contain a bright central galaxy ($M>>M_\mathrm{min}$). The satellite occupation function is 
\be 
\langle N_{\mathrm{sat}}\rangle = \frac{1}{2}\left[1+\mbox{erf}\left(\frac{\log M-\log 2M_{\mathrm{min}}}{\sigma_{\log M}}\right)\right]\left(\frac{M}{M_{\mathrm{sat}}}\right)^{\alpha_{\mathrm{sat}}}
\label{eq:hod_sat}. 
\ee
 It has a cut-off of the same form as the central occupation with a transition mass that is twice higher than that of the central galaxies. This is made to prevent halos having a low probability of hosting a central galaxy to contain satellite galaxies. The number of satellite galaxies grows with a slope \alphas. Making use of the HOD, the angular power spectrum and bispectrum of CIB anisotropies depend on four halo model parameters : \alphas, \Mmin, \Msat, and \sigmalogm.\\ 
The halo model formalism has only been applied to DSFG in the last few years but it has been extensively and successfully used to reproduce the clustering of optically selected galaxies and studied on simulations \citep{2010AA...518L..22C,2010ApJ...709...67T,2012A&A...542A...5C}. Even if numerical hydrodynamical simulations as well as semi-analytic models predict that \alphas~equals one \citep{2004ApJ...609...35K,2004MNRAS.355..819G,2005ApJ...633..791Z,2006MNRAS.365...11C}, measurements are not so unambiguous. \alphas~tends to be higher than one for DSFG but no convergence on its value is achieved yet.  \citet{2010AA...518L..22C} computed the correlation function of the sources detected in {\tt Herschel/SPIRE} data at 250, 350, and 500 \um~and \alphas~was found to range between 1.3 and $>1.8$. Concerning unresolved sources, results from {\tt Herschel/SPIRE} \citep{2011Natur.470..510A} and {\tt Planck} \citep{2011A&A...536A..18P} lead to values between 0.96 and 1.8 depending on the wavelength.  More recently, \citet{2012MNRAS.422.1324X} succeeded in fitting {\tt Planck,Herschel/SPIRE} and {\tt the South Pole Telescope} data from 250 \um~to 1.3 mm by one single model, and obtained \alphas=1.81$\pm$0.04. \\
 Both \Mmin~and \Msat~are hardly constrained separately as they are highly degenerate when dealing with CIB anisotropies power spectra \citep{2012A&A...537A.137P}. Therefore, one parameter is usually fixed whereas the other is derived from the data \citep{2011A&A...536A..18P,2012MNRAS.422.1324X} as inspired from theoretical studies and semi-analytic models which expect \Msat=10-25\Mmin. \citet{2011A&A...536A..18P} assumed \Msat=3.3 \Mmin~and found values of $\log (M_\mathrm{min}/ M_\odot)$~between 11.82 and 12.50.

\subsection{The emissivities} 
We use three models of evolution of DSFG throughout this study and compare their output emissivities. These three models are based on different philosophies (semi-analytical, backward evolution)  and all reproduce well current measurements but do predict different redshift evolution.

\subsubsection{Models of evolution of galaxies}\label{sect:model_evolution} 
Several models of the evolution of DSFG have been proposed in the literature \citep{2011ApJ...742...24L,2011MNRAS.416...70G,2011A&A...529A...4B,2012ApJ...757L..23B}. In the present study we focus on three models \citep{2011A&A...529A...4B,2012ApJ...757L..23B,2007MNRAS.377.1557N} all reproducing well observations (differential number counts, luminosity functions at several redshifts and differential number counts in redshift bins from {\tt Herschel}). Their differences allow us to explore and compare the effects of the evolution of DSFG, in particular, on the bispectrum of the CIB anisotropies.\\ 
 First, we consider an updated version of the model of \citet{2007MNRAS.377.1557N} (hereafter {\sc model1}) that has been recently used by \citet{2012MNRAS.422.1324X} to model the power spectrum of the CIB anisotropies. This model uses two populations of galaxies, spirals at $z<1.5$ and high-$z$ sub-mm galaxies at $z>1$. The latter are considered to be massive proto-spheroidal galaxies processing most of their stellar mass. Furthermore, the model assumes that star formation is triggered by the dark matter halo collapse/merger and is then controlled by self-regulated baryonic processes such as AGN feedback.\\  The second model (hereafter {\sc model2}), proposed by \citet{2011A&A...529A...4B}, is a backward evolution model. It relies on a parametric luminosity function (LF) and a library of spectral energy distributions (SED) templates. The LF is taken to be a power law for low luminosities and a Gaussian for high luminosities. \citet{2011A&A...529A...4B} used the SED library of \citet{2004ApJS..154..112L} which contains two different populations of galaxies: star-forming and late-type. The former emit over 95\% of their energy in the IR while the latter emit half or less of their energy in the IR. {\sc model2} is described by thirteen free parameters. Best-fit parameters are computed using Monte-Carlo Markov Chains on available differential number counts and luminosity functions on a large range of IR wavelengths.\\ 
 The third model (hereafter {\sc model3}), from \citet{2012ApJ...757L..23B}, is an empirical model in which galaxies are split between main sequence (MS) galaxies and starburst (SB) ones. MS galaxies account for 85\% of the star-formation density of the Universe at $z<2$. The 15\% left are due to SB galaxies which have high specific star-formation rates (sSFR), likely because of recent major mergers. In their model, \citet{2012ApJ...757L..23B} assume an SED that gets warmer with redshift for MS galaxies and a non-evolving SED for SB galaxies. Both SEDs are based on {\tt Herschel} observations. The sSFR distribution is thus decomposed in the two modes, SB and MS. Following \citet{2012ApJ...747L..31S}, the sSFR of MS galaxies vary with redshift and stellar mass whereas it remains constant for SB galaxies. The stellar mass function changes with the redshift as well. In addition, the model includes dust attenuation, AGN contribution and, magnification by strong lensing.

\begin{figure}\centering \includegraphics[scale=0.5]{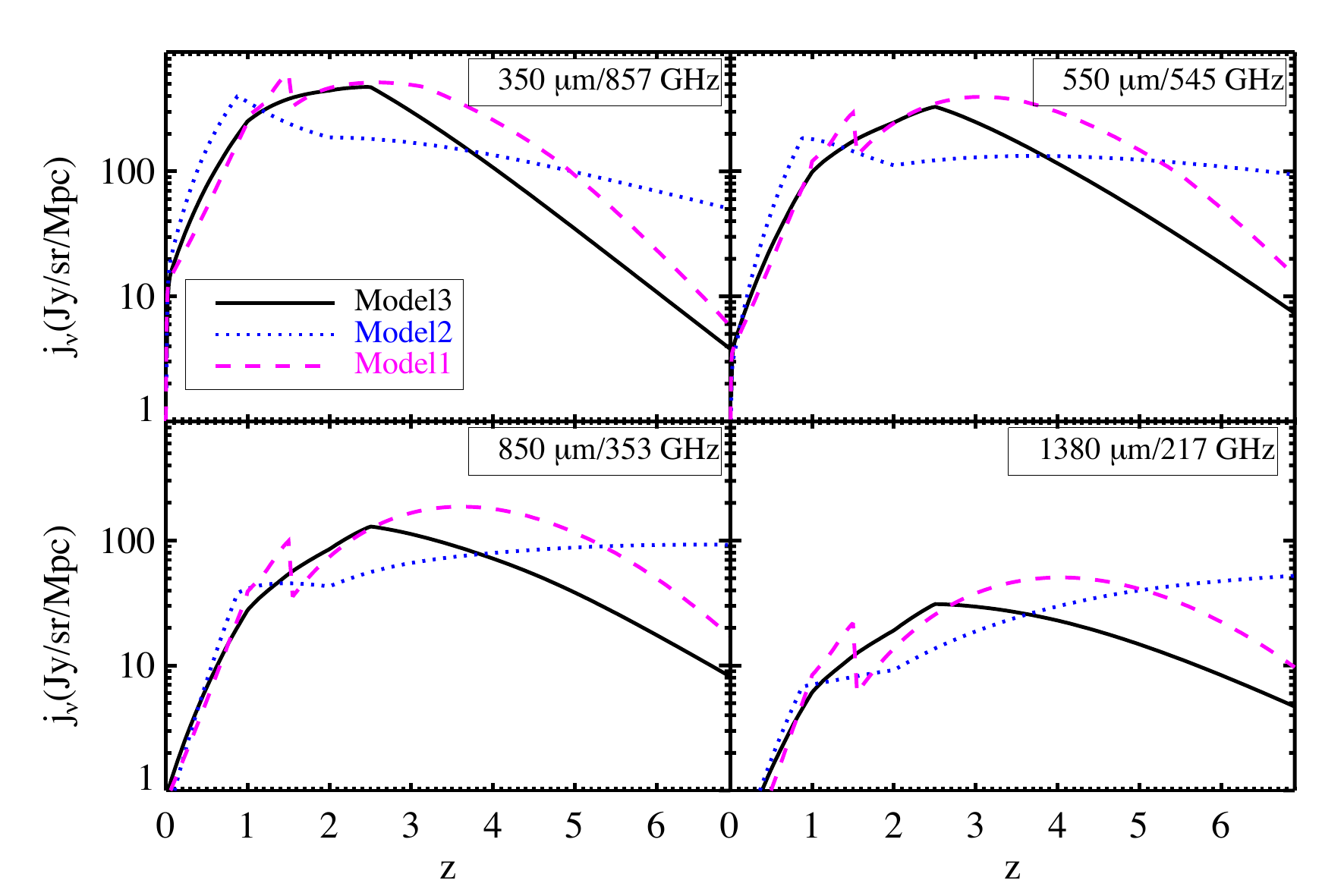} \caption{Emissivities for the three models of evolution of galaxies at long wavelengths used in the present study.} \label{fig:plot_jd} \end{figure} 

\subsubsection{The emissivities} 
We derive the mean emissivities from the differential number counts as follows: 
\be 
\bar{j}_\lambda(z)=(1+z)\int_0^{S_{\mathrm{cut}}} S\frac{\dd^2 N}{\dd S\dd z}\dd S. 
\ee 
We display the emissivities at long wavelengths for the three models of evolution of galaxies in Fig. \ref{fig:plot_jd}. The cut-offs at $z\sim 1$ and $z=2$ of {\sc model2} are due to the parametrisation of the luminosity function which has two breaks at these redshifts. The sharp break at $z=1.5$ of {\sc model1} is the result of the mix of two populations. The peak is caused by the contribution of starbursts galaxies.\\ 
The emissivities peak at different redshifts. The peak of the emissivity is shifted from $z\sim1$ for {\sc model2} to $z\sim2.5$ for {\sc model3}, in better agreement with the peak of star formation rate in the Universe at $z\sim2-3$ \citep{2012ApJ...754...83B}.  {\sc model3} also displays a peak at $z\sim2$ at 350 \um~that shifts up to $z\sim4$. The three emissivities display a similar shape and amplitude up to $z\sim1.5$. However, their behaviour is different at higher redshift. {\sc model1} and {\sc model3} emissivities peak at $z\sim2-4$ for all the wavelengths whereas those of {\sc model2} are either decreasing (350 \um) constant (550 and 850 \um) or increasing (1380 \um). At very high redhifts ($z>4$), {\sc model1} and {\sc model3} emissivities decrease whereas those of {\sc model2} stay constant or increase. The behaviours of {\sc model1} and {\sc model3} are more in line with our recent knowledge of the evolution of the SFR with redshift \citep{2011ApJ...737...90B}. We therefore adopt {\sc model3} as a baseline for the present study and we use the other models of evolution of galaxies to explore the range of possibilities for the bispectrum.


\section{Predictions of the bispectrum and dependencies on the models of evolution of galaxies}\label{par:pred+gal+hod} 
\begin{figure*}\centering 
\begin{tabular}{cc} 
\includegraphics[scale=0.45]{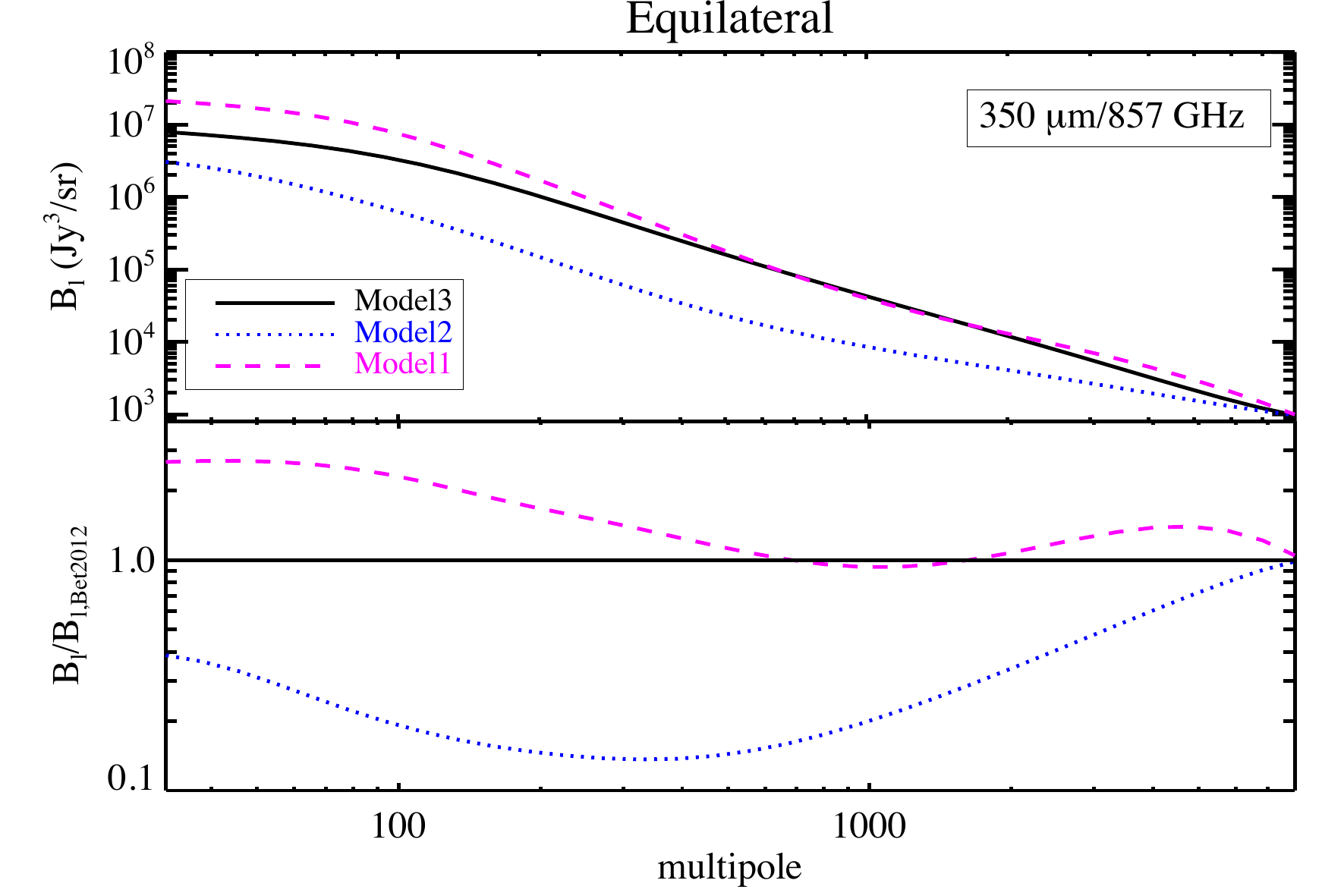} & \includegraphics[scale=0.45]{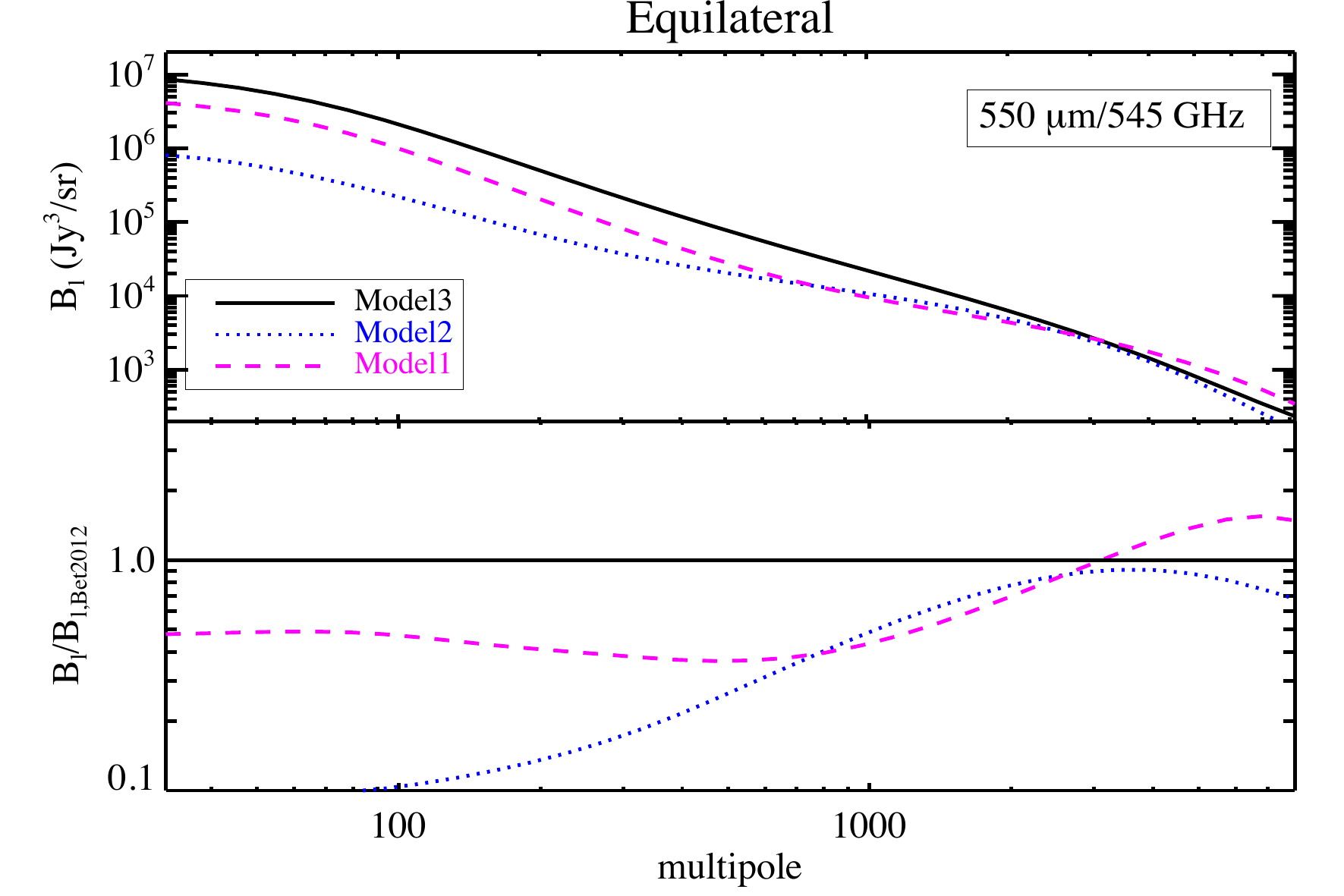} \\ \includegraphics[scale=0.45]{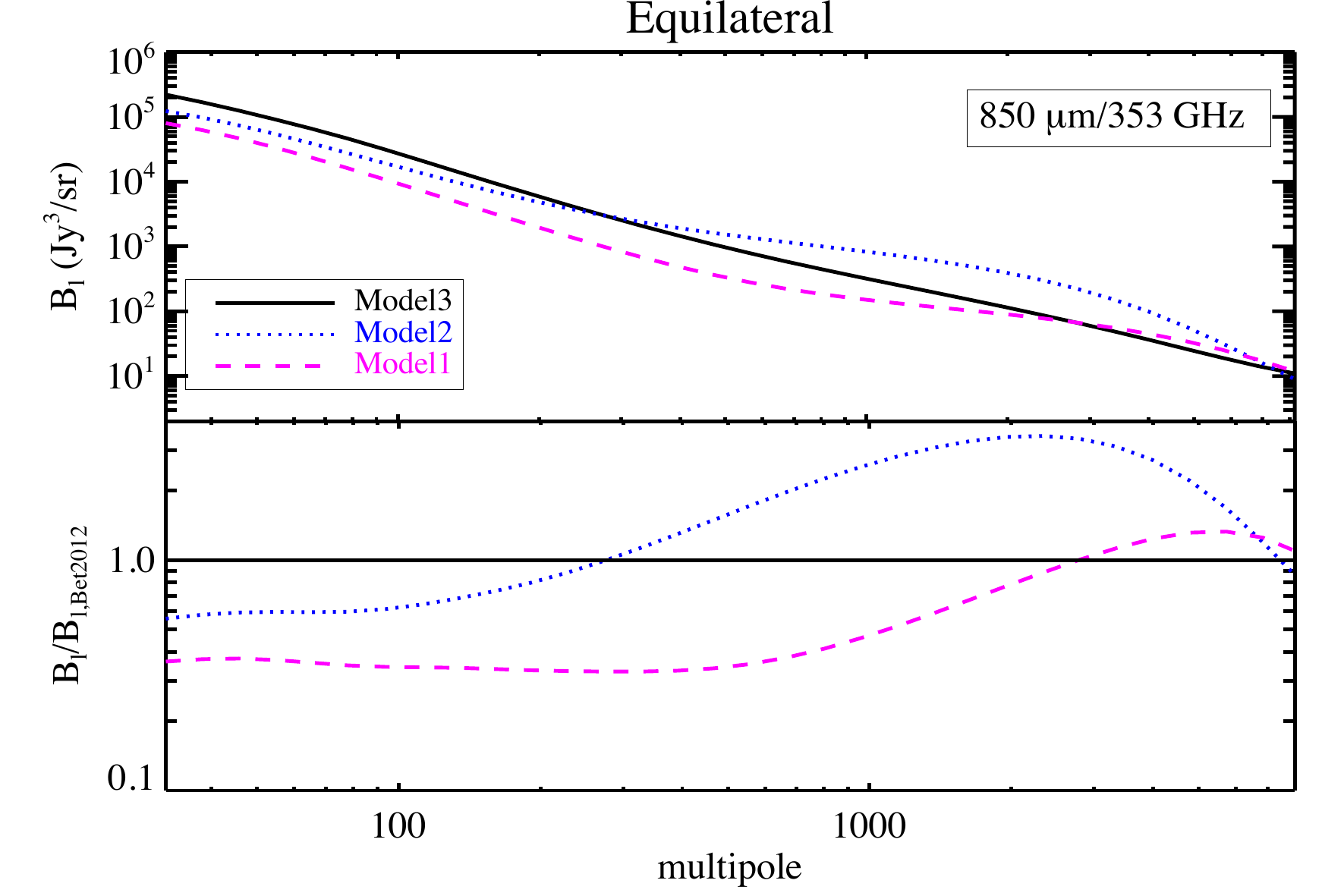} & \includegraphics[scale=0.45]{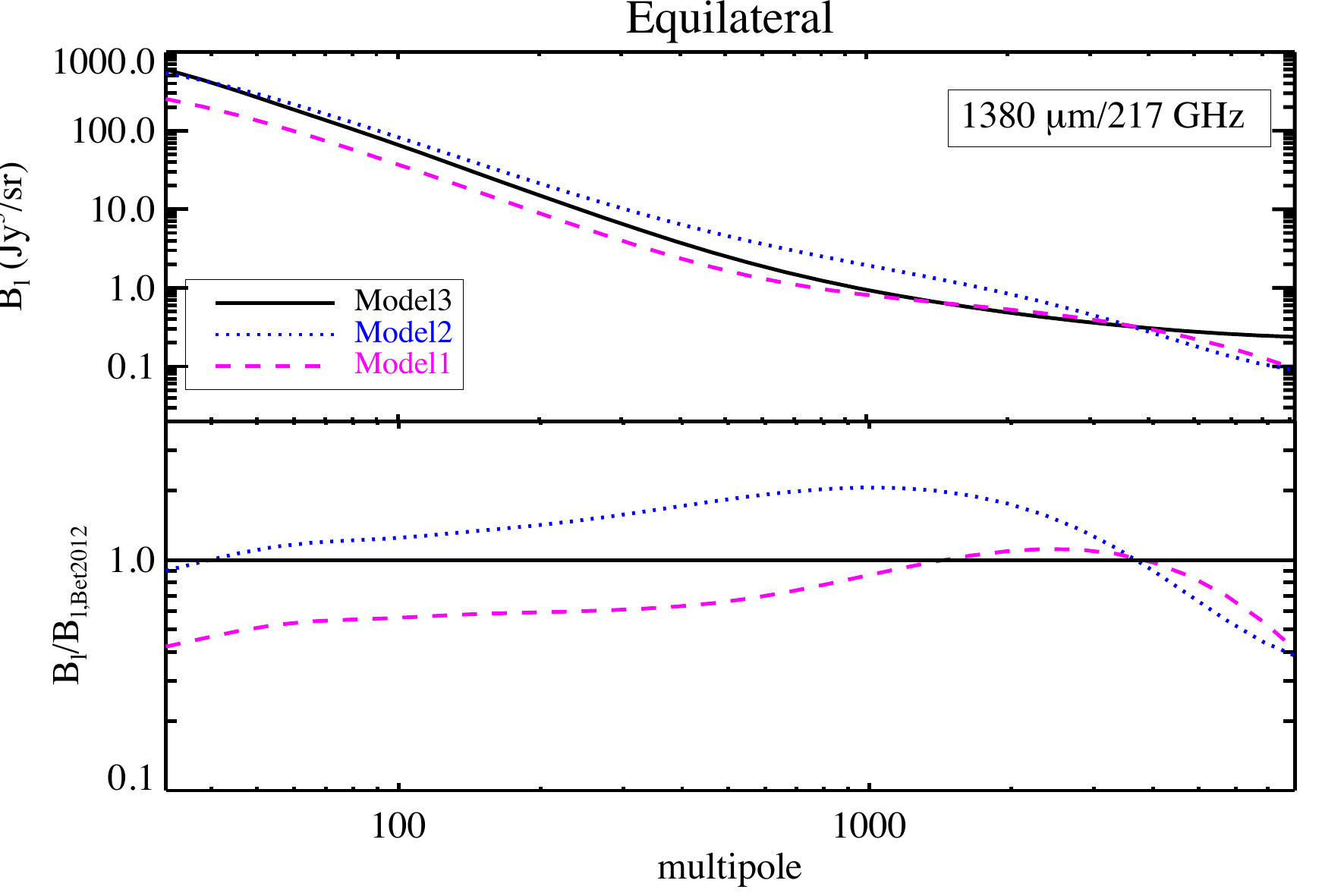} 
\end{tabular} 
\caption{Top panels: Equilateral bispectra with the three models of evolution of galaxies. Lower panels: ratios between bispectra compared to the one computed with {\sc model3}.} \label{fig:compare_bl_jnu_equi}
 \end{figure*}

\begin{figure*}\centering 
\begin{tabular}{cc}
 \includegraphics[scale=0.45]{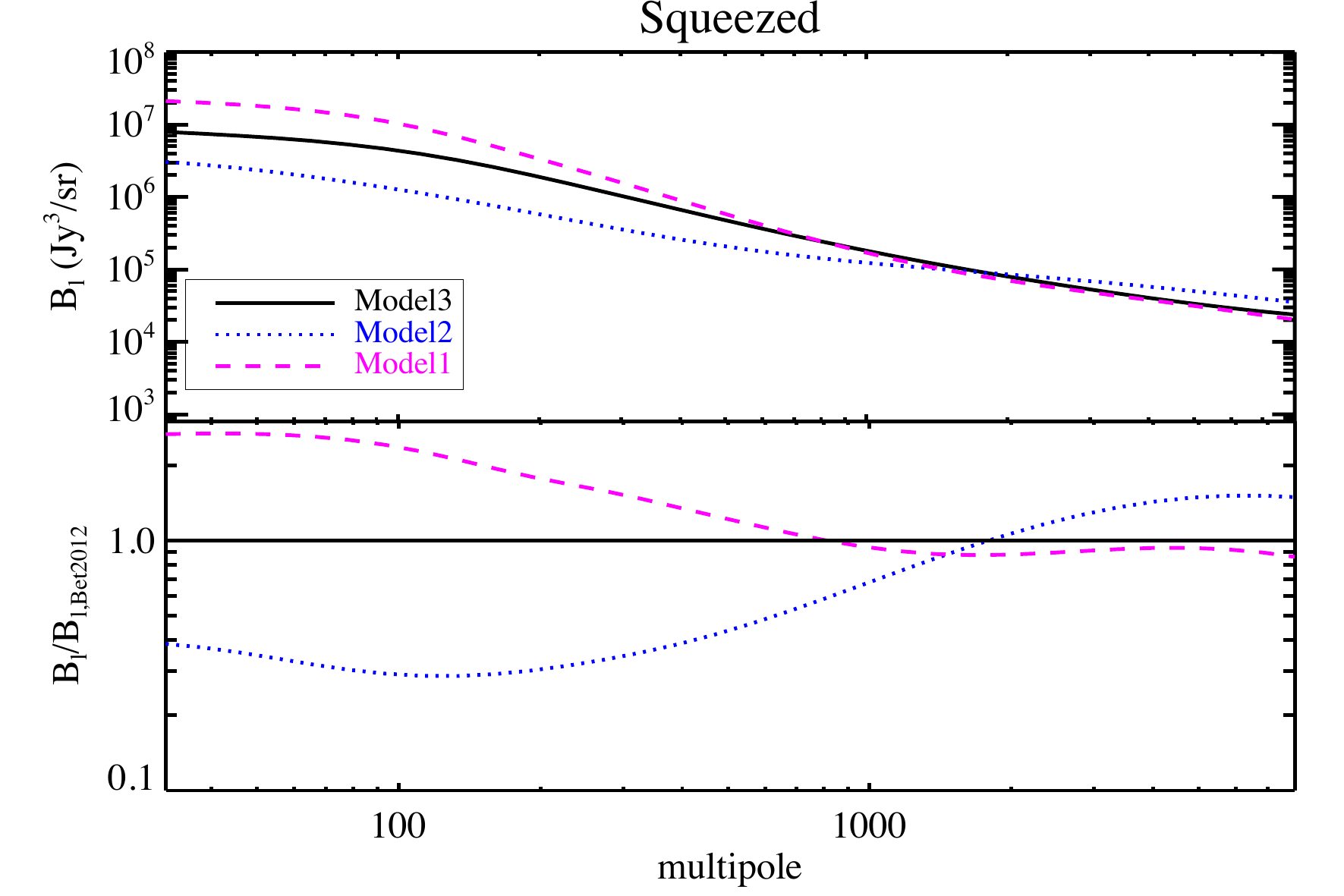} & \includegraphics[scale=0.45]{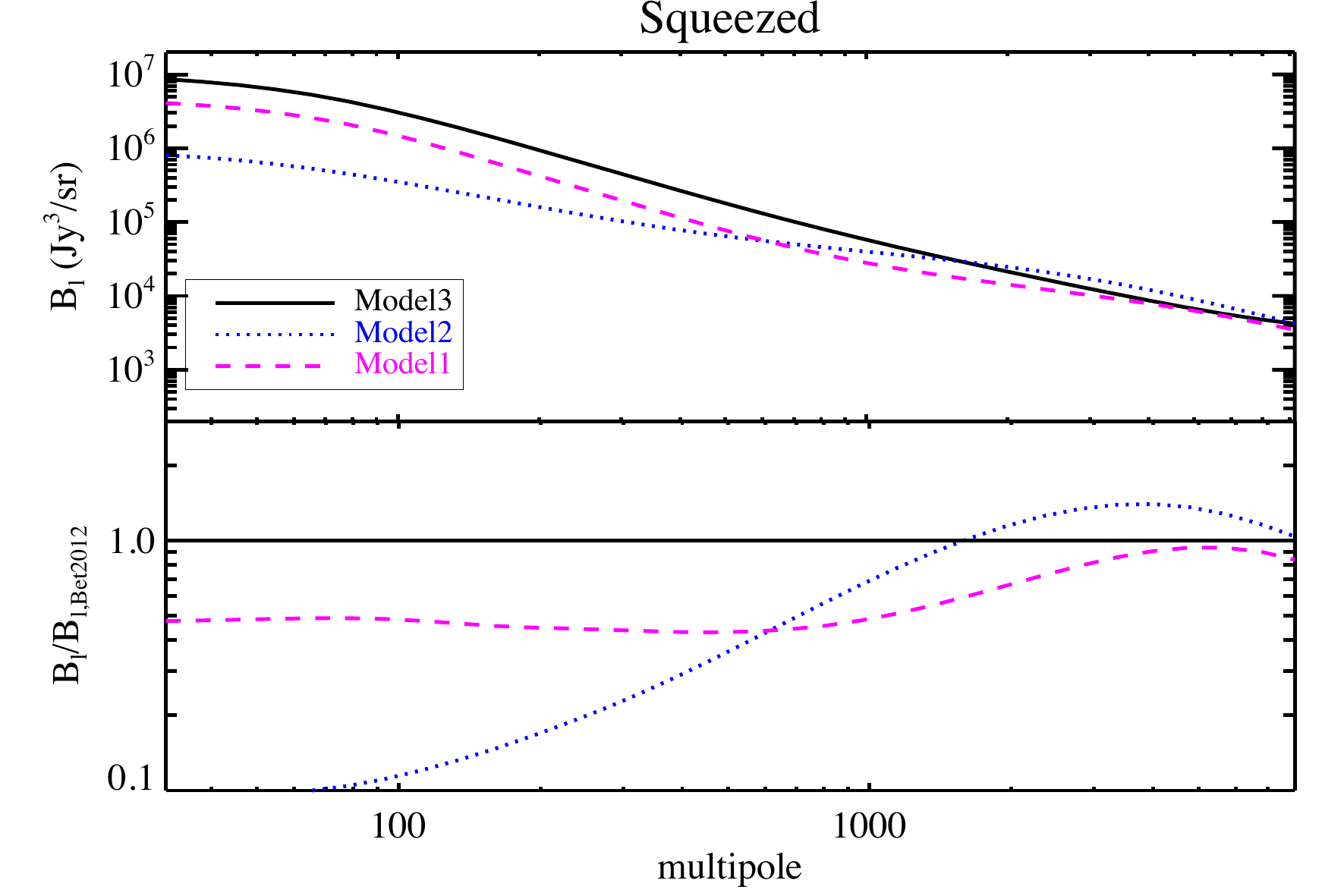} \\ \includegraphics[scale=0.45]{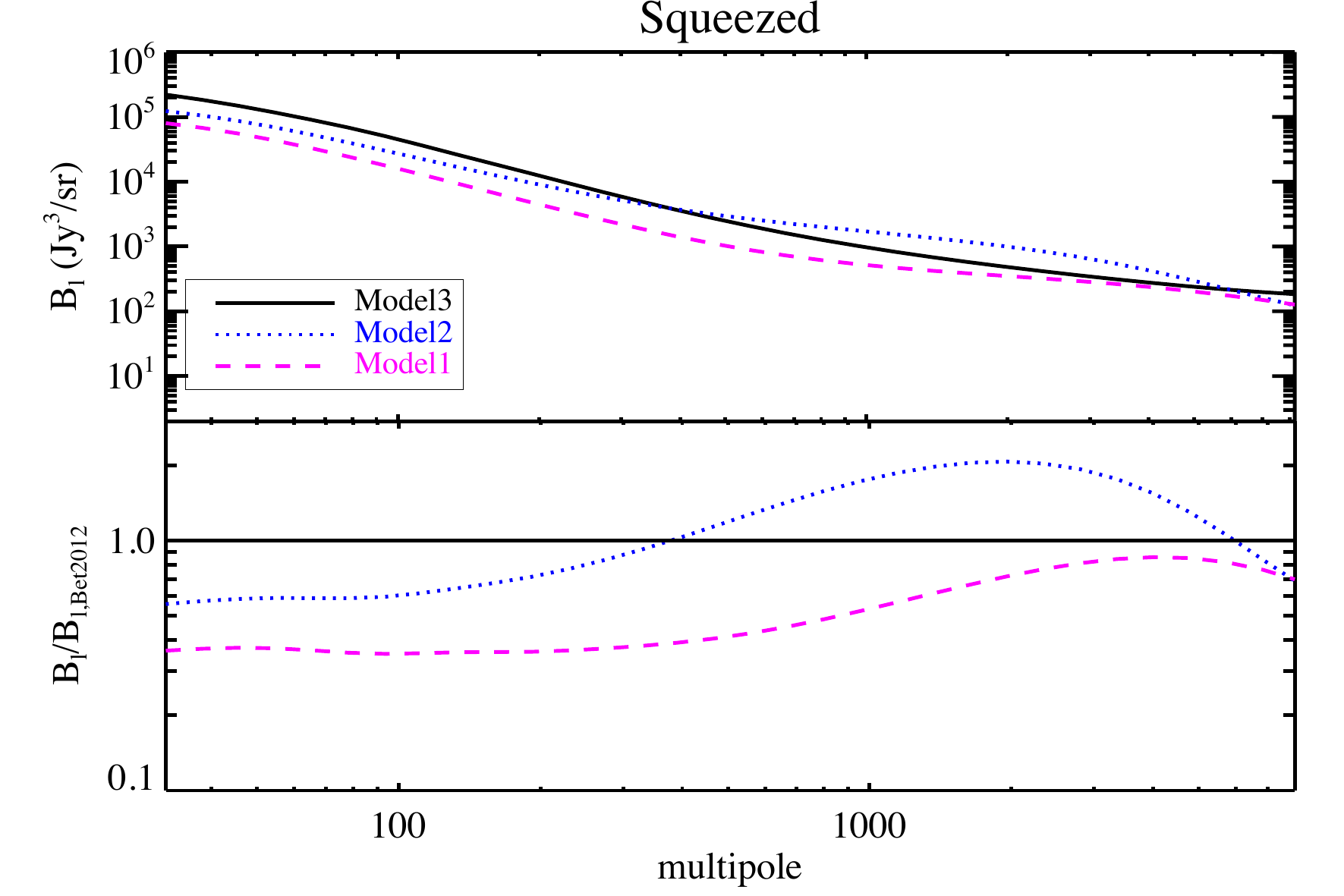} & \includegraphics[scale=0.45]{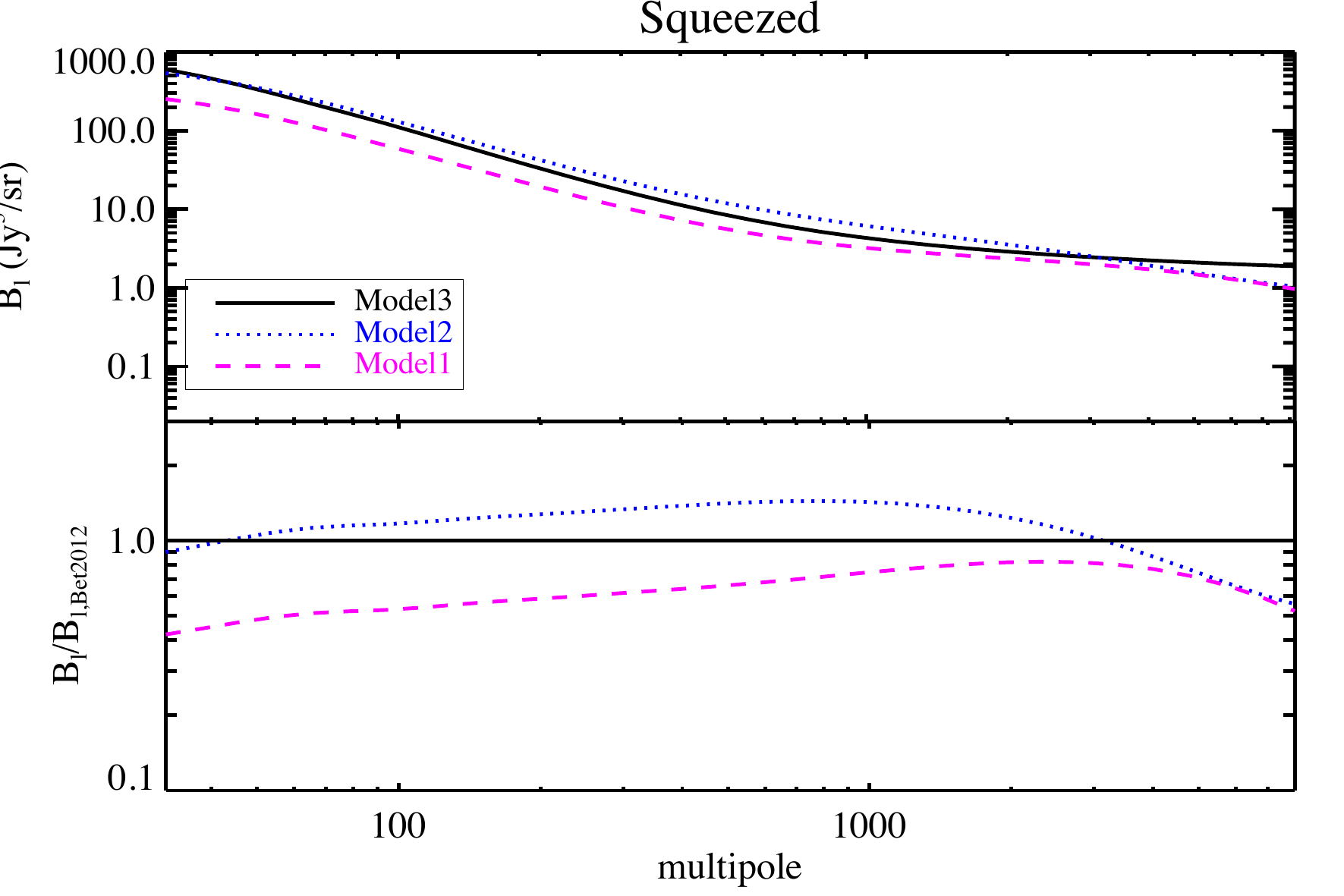} 
\end{tabular}
 \caption{Top panels: squeezed bispectra computed with the three models of evolution of galaxies at long wavelengths. Low panels: ratios between {\sc model2} and {\sc model1} bispectra compared to the one computed with {\sc model3}.} \label{fig:compare_bl_jnu_squeezed} 
\end{figure*}

We now review how the predictions of the bispectrum are computed and to what extent they depend on the emissivities. The latter are already known to have a strong influence on the shape and on the amplitude of the power spectrum \citep{2012A&A...537A.137P}.\\
\citet{2011A&A...536A..18P} and \citet{2012MNRAS.422.1324X} derived sets of HOD parameters associated to their respective models of galaxies using the \citet{2008ApJ...688..709T} mass function. However the second order bias, which is a key ingredient to compute the bispectrum, is not available for the \citet{2008ApJ...688..709T} mass function. We therefore make use of the \citet{1999MNRAS.308..119S} mass function and its associated first and second order biases. Consequently, we carry our fits of {\tt Planck} power spectra for each wavelength and galaxy model under consideration. To this end, we fit the HOD parameters \alphas~and \Mmin, fixing \Msat=10 \Mmin~and \sigmalogm~=~0.65, at each wavelength, as HOD parameters are degenerate with the emissivity model. The fits are satisfactory, with reduced $\chi^2$ ranging between 1 and 2. Best-fits values are given in Table. \ref{tab:best_fit}. Making use of the obtained best-fit HOD parameters, we compute the equilateral and squeezed bispectra at the four {\tt Planck} wavelengths, 350, 550, 850, and 1380 \um. They are shown in Figs. \ref{fig:compare_bl_jnu_equi} and \ref{fig:compare_bl_jnu_squeezed}. Lower panels display the ratio between bispectra with respect to that obtained with {\sc model3}. The difference does not depend much on the configuration but mainly on the wavelength and the scale. The ratio at `low' wavelength between the bispectra of {\sc model2} and {\sc model3} is mainly lower than one. Regardless of the configuration, this ratio increases, and exceeds unity, with increasing wavelengths. The ratio between {\sc model1} and {\sc model3} stays below one between 550 and 1380 \um. At 350 \um, this ratio is above one and reaches three. It depends on the scale. Regardless of the configuration and the wavelength, there is always a scale range where bispectra differ by at least a factor two. The main differences are at low multipoles ($\ell<300$). For instance, at $\ell=100$, it varies between a factor nine for the equilateral at 550 \um~and a factor 1.1 at 1380 \um~for the squeezed configuration. At higher multipoles ($\ell>1000$), the differences are smaller. At $\ell=4000$, they are mostly around a factor 1-1.5 even if it reaches 2.7 for the equilateral at 850~\um. These differences arise from the set of parameters of the HOD as well as from the model of galaxies.\\ 
Such differences in the predictions lead to a difference of the level of non-Gaussianity in the CIB anisotropies. By computing the ratio of the total bispectra for each model of galaxies we can get an insight on that level. {\sc model2} predicts more non-Gaussianity than {\sc model3} up to 850 \um. The relative level predicted by {\sc model1} varies with respect to the wavelength up to 850 \um. At 1380 \um, the amount of non-Gaussianity is similar with the three models of galaxies. \citet{2013arXiv1303.3535C} measured the amplitude of the bispectrum of the CIB at 220 GHz and $\ell=2000$. They found 0.87$\pm0.29\pm0.25$ $\mu$K$^3$/sr (the first error is statistical and the second one is systematical) whereas we predict 3.86 $\mu$K$^3$/sr for {\sc model3}. One explanation for this disagreement could be that our predictions have been calibrated on power spectra with $\ell\in$ [200,2000] whereas {\sc SPT} is dedicated to higher multipoles.\\ 
Such behaviours show that fitting simultaneously power spectra and bispectra will lead to the exclusion of some models of evolution of galaxies that predict too much or not enough non-Gaussianity. We also note that the main differences are displayed at small multipoles ($\ell<100$), therefore large CIB maps will be needed to be able to measure accurately the bispectrum on these scales. However, these scales are the most affected by Galactic dust and, at 1380 \um~by CMB anisotropies \citep{2013arXiv1309.0382P}. Nevertheless, we can expect that the combination of power spectra and bispectra measurements will allow to disentangle between models of evolution of galaxies.


\begin{table*}\centering 
\begin{tabular}{cccccc}
 \hline\hline 
Wavelength                  &Frequency               & parameters &{\sc model1} &{\sc model2} &{\sc model3}\\
 \um                            & GHz &&&\\  
\hline 
\multirow{2}{*} {350}   &\multirow{2}{*}{857} & \alphas                 & 1.8   & 1.1   &1.7\\  
                                                                    && $\log_{10}$\Mmin & 12.8 & 12.6 &12.8\\ 
\hline 
\multirow{2}{*} {550} &\multirow{2}{*}{545}   & \alphas                 & 1.7 & 1.7 &1.9\\  
& & $\log_{10}$\Mmin & 12.2 & 12.8 &12.6\\ 
\hline 
\multirow{2}{*} {850} &\multirow{2}{*}{353} & \alphas & 1.5 & 1.9 &1.7\\  
& & $\log_{10}$\Mmin & 11.6 & 12.6 &12.2\\ 
\hline \multirow{2}{*} {1380} &\multirow{2}{*}{217} & \alphas & 1.4 & 1.6 &1.5\\  
& & $\log_{10}$\Mmin & 11.2 & 12.2 &12\\ 
\hline 
\end{tabular} 
\caption{Best-fits of the HOD parameters for {\sc Planck} power spectra used to compute predictions of the bispectrum. We fixed \Msat = 10\Mmin~following \citet{2011A&A...536A..18P}.} 
\label{tab:best_fit}
 \end{table*}


\section{Redshift and halo mass contributions}\label{par:zm_distrib} 
We investigate the redshift contributions of the bispectrum compared to those of the power spectrum and eventually we explore the halo mass contribution as a function of the redshift.

\subsection{Redshift contributions}\label{par:redshift_distrib} 
\begin{figure*}\centering 
\includegraphics[scale=0.65]{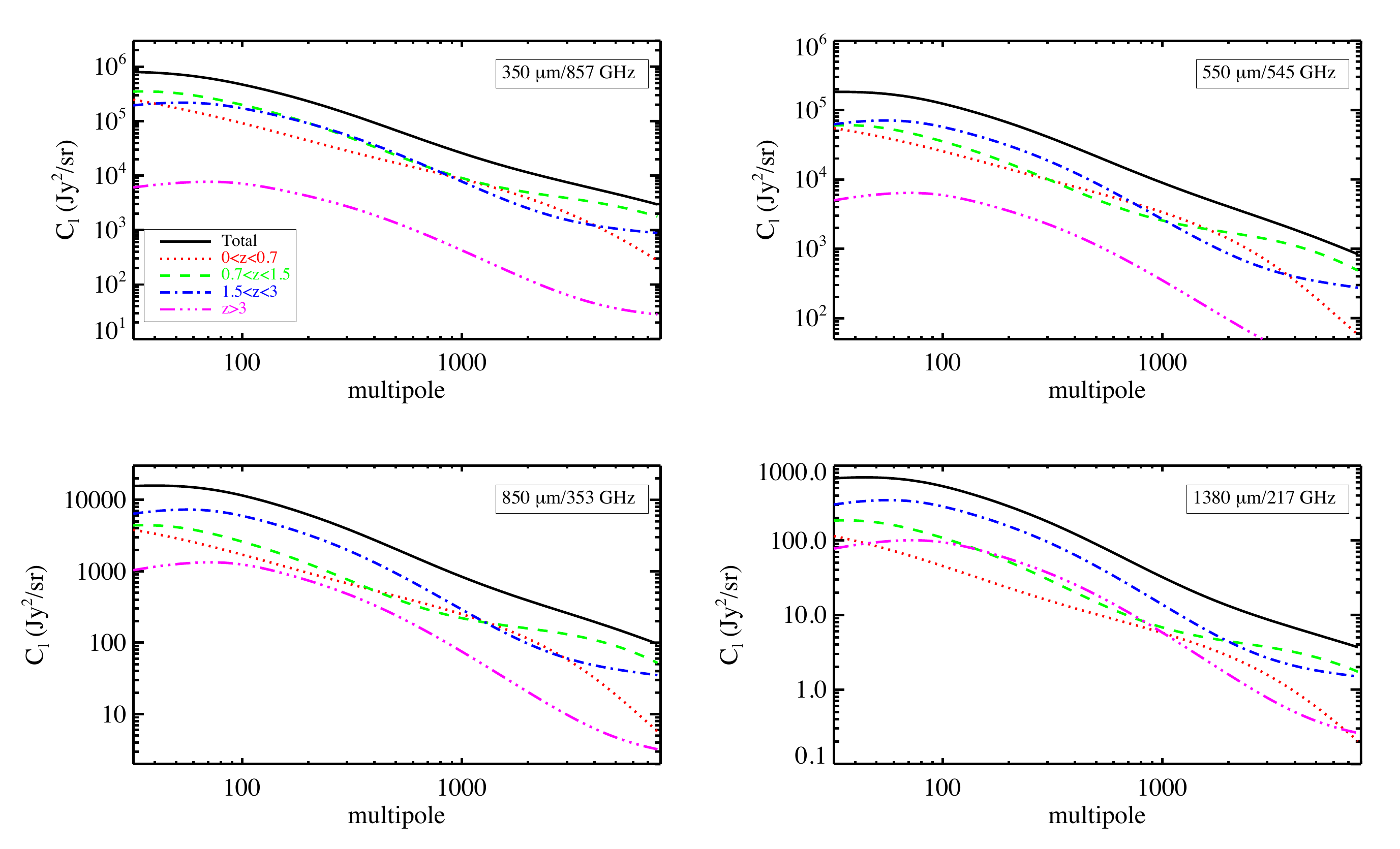} 
\caption{Redshift contributions of the power spectra coming from {\sc model3}.} 
\label{fig:compare_cl_jnu_planck_z} 
\end{figure*} 
\begin{figure*} 
\includegraphics[scale=0.69]{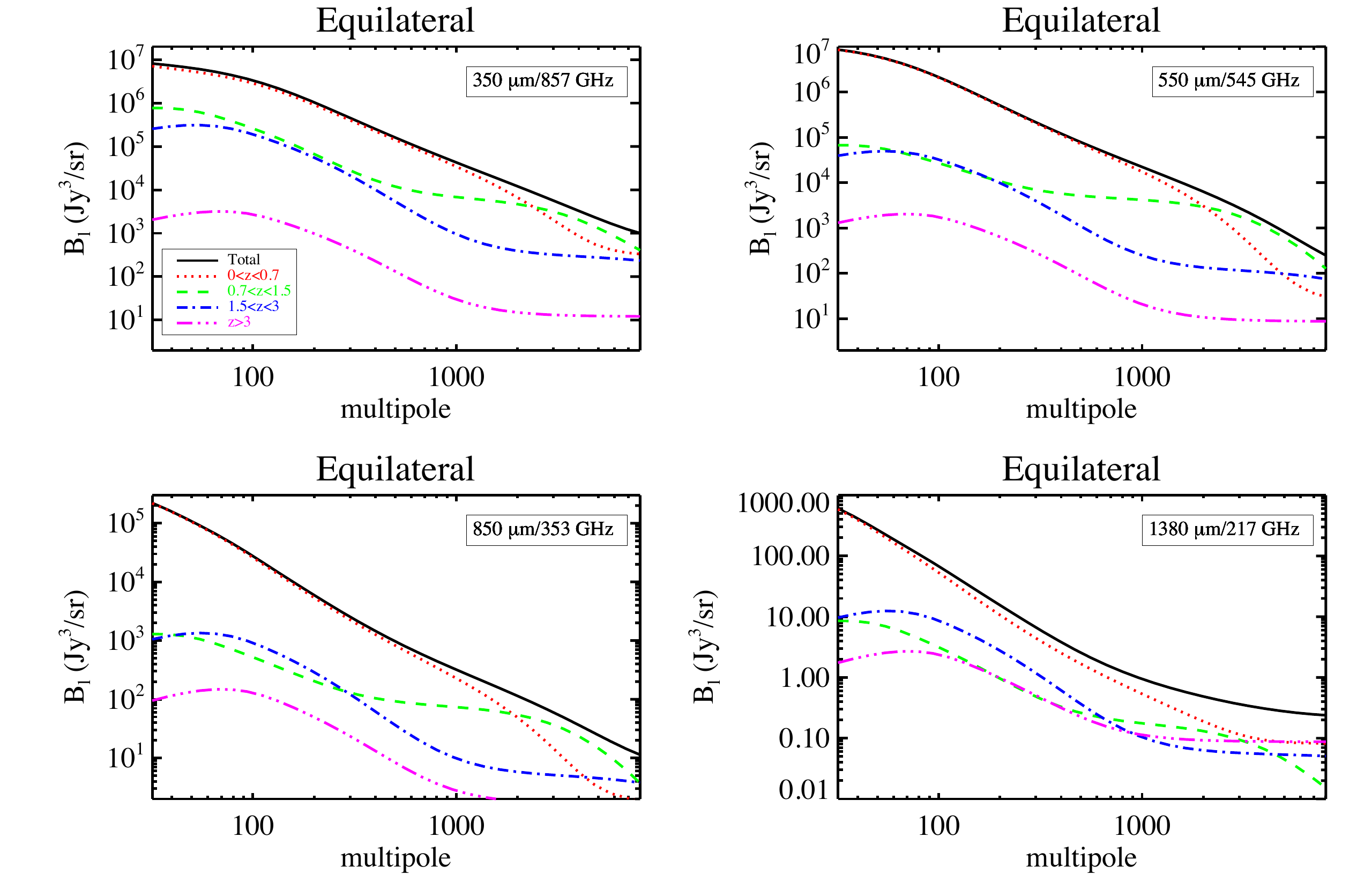} 
\caption{Redshift contributions of the equilateral bispectra coming from {\sc model3}.}
 \label{fig:compare_bl_jnu_planck_z_equi_bet2012} 
\end{figure*}
 \begin{figure*}\centering 
\includegraphics[scale=0.69]{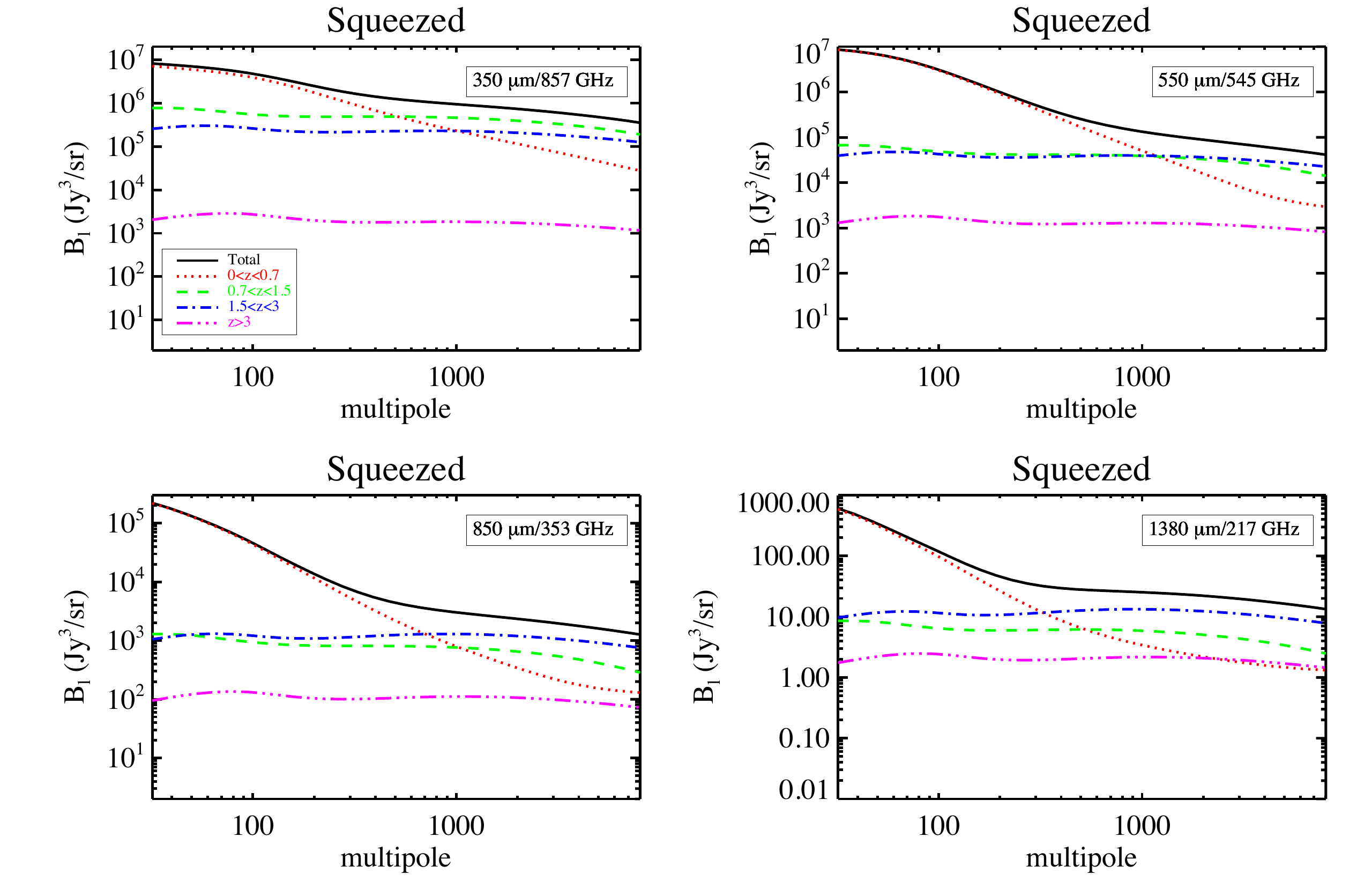} 
\caption{Redshift contributions of the squeezed bispectra coming from {\sc model3}.}
 \label{fig:compare_bl_jnu_planck_z_squeezed_bet2012} 
\end{figure*}

\begin{figure*}\centering 
\includegraphics[scale=0.45]{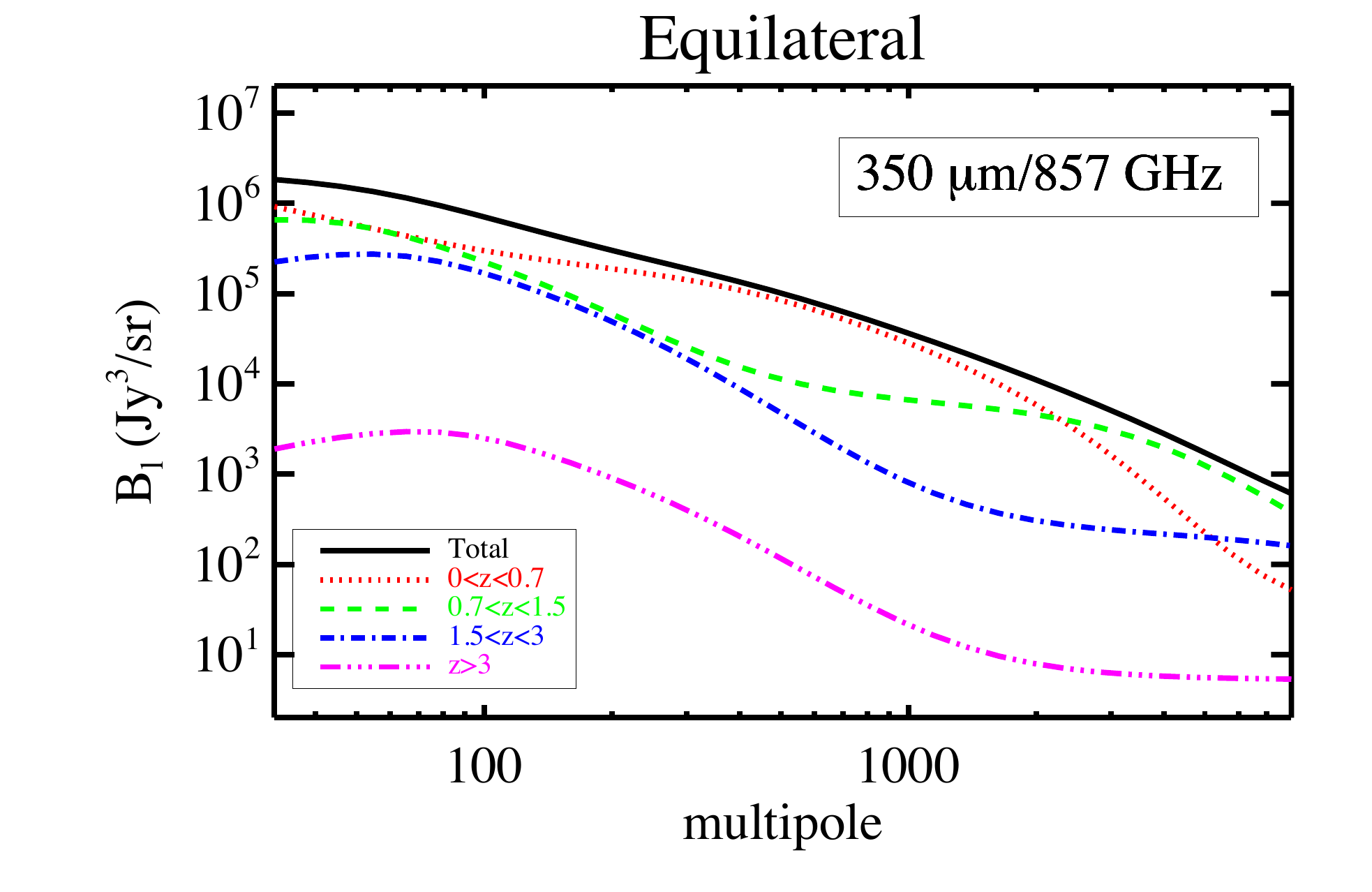}\includegraphics[scale=0.45]{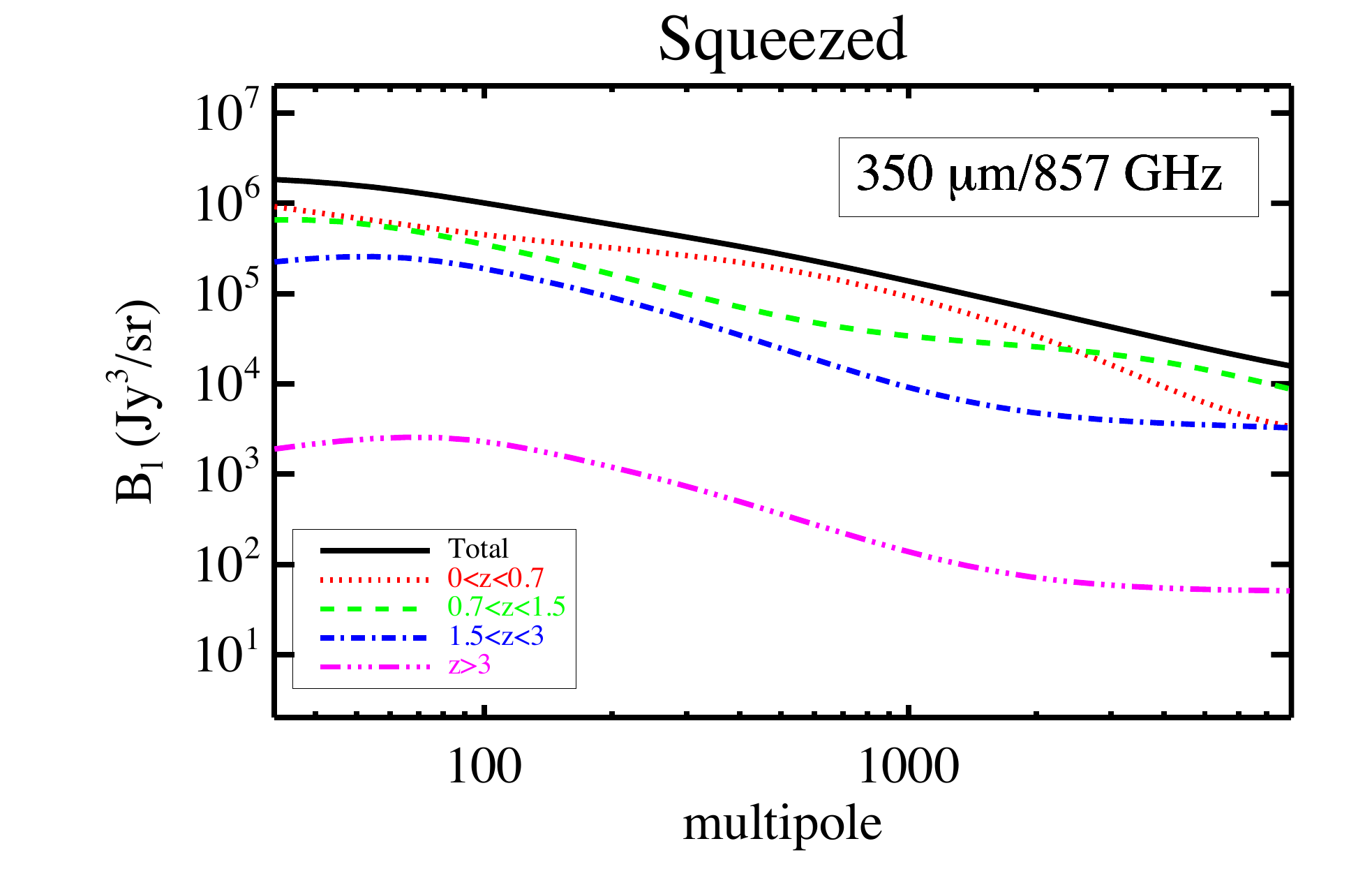} 
\caption{Redshift contributions of the equilateral and squeezed bispectra coming from {\sc model3} with low flux cuts.} 
\label{fig:contrib_z_bl_350_low_fc} 
\end{figure*}

\begin{figure*}\centering 
\includegraphics[scale=0.63]{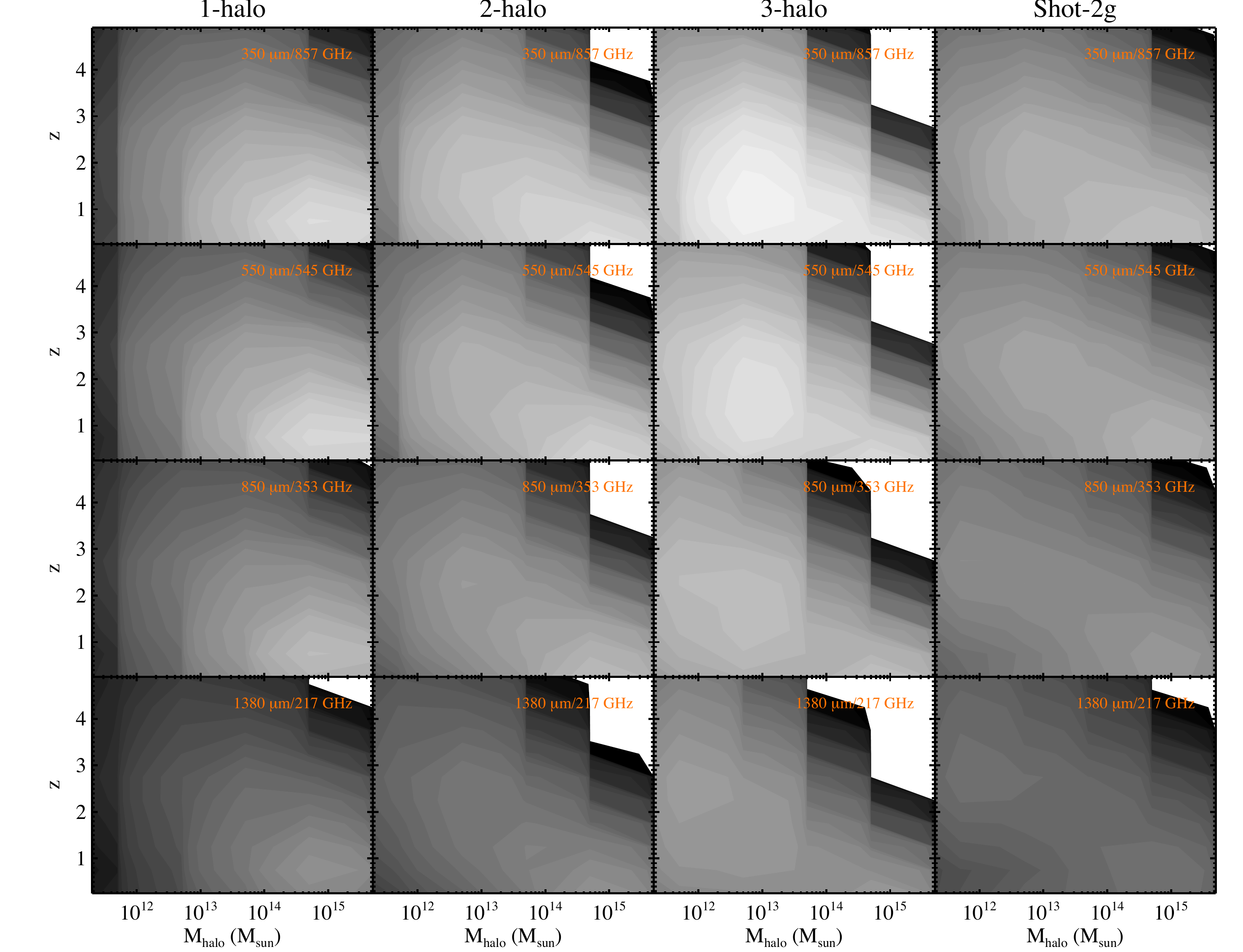} 
\caption{Halo mass contributions as a function of the redshift to the different terms of the equilateral bispectrum at several wavelengths. The light grey corresponds to the highest contribution. The step of color is logarithmic.} \label{fig:bl_planck_contrib_mz_equi} 
\end{figure*}
One of the main unknown of the CIB is its redshift distribution.  Recent measurements of differential number counts per redshift bins start to uncover its nature up to redshift 1.5 \citep{2011A&A...525A..52J,2012A&A...542A..58B}.\\  First we investigate the redshift distribution of the predictions of the bispectrum and compare them to those of the power spectrum.  We use the prediction coming from {\sc model3}.\\ 
The redshift contributions of the power spectrum are shown in Fig. \ref{fig:compare_cl_jnu_planck_z}. As the wavelength increases, the contribution of low to moderate redshift galaxies ($0<z<0.7$ and $0.7<z<1.5$) decreases while that of higher-redshift galaxies ($1.5<z<3$ and $z>3$) increases. Fig. \ref{fig:compare_bl_jnu_planck_z_equi_bet2012} and \ref{fig:compare_bl_jnu_planck_z_squeezed_bet2012} show the redshift contributions of, respectively, the equilateral and squeezed bispectra. In both case, we do recover the same trend observed for the power spectrum. The relative contribution of high redshift galaxies gets larger as the wavelength increases while that of low-redshift galaxies decreases. However, bispectra at each wavelength and in both configurations are dominated by the contribution of galaxies at low redshift ($0<z<0.7$), which is due to a projection effect ($1/r^4(z)$ factor in Eq.\ref{Eq:bisp_3Dtoang}). This domination does not arise on all scales, and the domination range depends on wavelength as well as configuration. Indeed, other redshift bins may dominate depending on scale wavelength and configuration.
configuration galaxies with $0<z<0.7$ dominate up to $\ell\sim300$. 
At higher multipoles, higher redshift galaxies dominate the bispectrum. The bin $0.7<z<1.5$ contributes the most at 350 \um~while the contribution of the range $1.5<z<3$ increases with the wavelength and dominates the squeezed bispectrum at $\lambda\geq$850 \um. Therefore it appears that the bispectrum will not give easily access to high redshift galaxies. Nevertheless, applying a flux cut to the emissivities is equivalent to removing low $z$ sources (see Sect. \ref{par:importance_fc}).

\subsection{Redshift and mass contributions simultaneously}

 We investigate the evolution of the mass contributions of each component with the redshift as shown in Fig. \ref{fig:bl_planck_contrib_mz_equi}. To this end, we consider the equilateral configuration at the multipole $\ell=1958$, $\ell=133$, $\ell=32$, and $\ell=1958$ for the 1-halo, 2-halo, 3-halo, and shot-2galaxies terms respectively. Indeed, these are the multipoles where the respective terms are important.\\
 The 1-halo term is dominated by galaxies hosted by high mass halos ($10^{14}$\Msun$<$\Mhalo$<10^{15}$\Msun) at $z\sim1$. As the redshift increases, the dominating mass range shifts towards lower halo mass, $\sim10^{13}$\Msun.\\
 The 2-halo term is also dominated by galaxies that are in halos of $10^{15}$\Msun~but at lower redshift, $z\sim0.5$. The dominating halo mass range shifts to lower mass with increasing redshift, up to $10^{13}$\Msun~at 350 \um~and up to $10^{11-12}$\Msun~at 1380 \um.\\
 Concerning the 3-halo term, halos from a large mass range ($10^{12}$\Msun$<$\Mhalo$<10^{15}$\Msun) as well as from a large redshift range ($0<z<2$) contribute equally at 350 \um.  As the redshift increases the mass range gets smaller and the dominating halo mass tends to $\sim10^{13}$\Msun. This trend is observed at the other wavelengths but the dominating mass at redshift higher than 2 gets even smaller, $10^{12}$\Msun~at 1380 \um. We can also notice that galaxies in halos with $10^{11}$\Msun$<$\Mhalo$<10^{14}$\Msun~from redshift 0.5 to 3 contribute equally at 1380 \um, while the mass range gets thinner towards high redshifts (and centered around $10^{12}$ \Msun). Indeed as the redshift increases the cut-off of the halo mass function goes to smaller halo masses.\\
 Lastly, the 2-galaxies shot noise is dominated by galaxies in massive halos of $10^{14}$\Msun$<$\Mhalo$<10^{15}$\Msun~at 350 \um. As the wavelength gets longer, the redshift/mass domination spreads to higher redshifts and to an intermediate halo mass, $10^{12}$\Msun~at 1380~\um.\\
 We do recover that higher redshift galaxies contribute relatively more at long wavelengths than at short ones and that each term is sensitive to a different mass range.


\section{The importance of the emissivities flux cuts}\label{par:importance_fc}

\begin{figure*}\centering \includegraphics[scale=0.78]{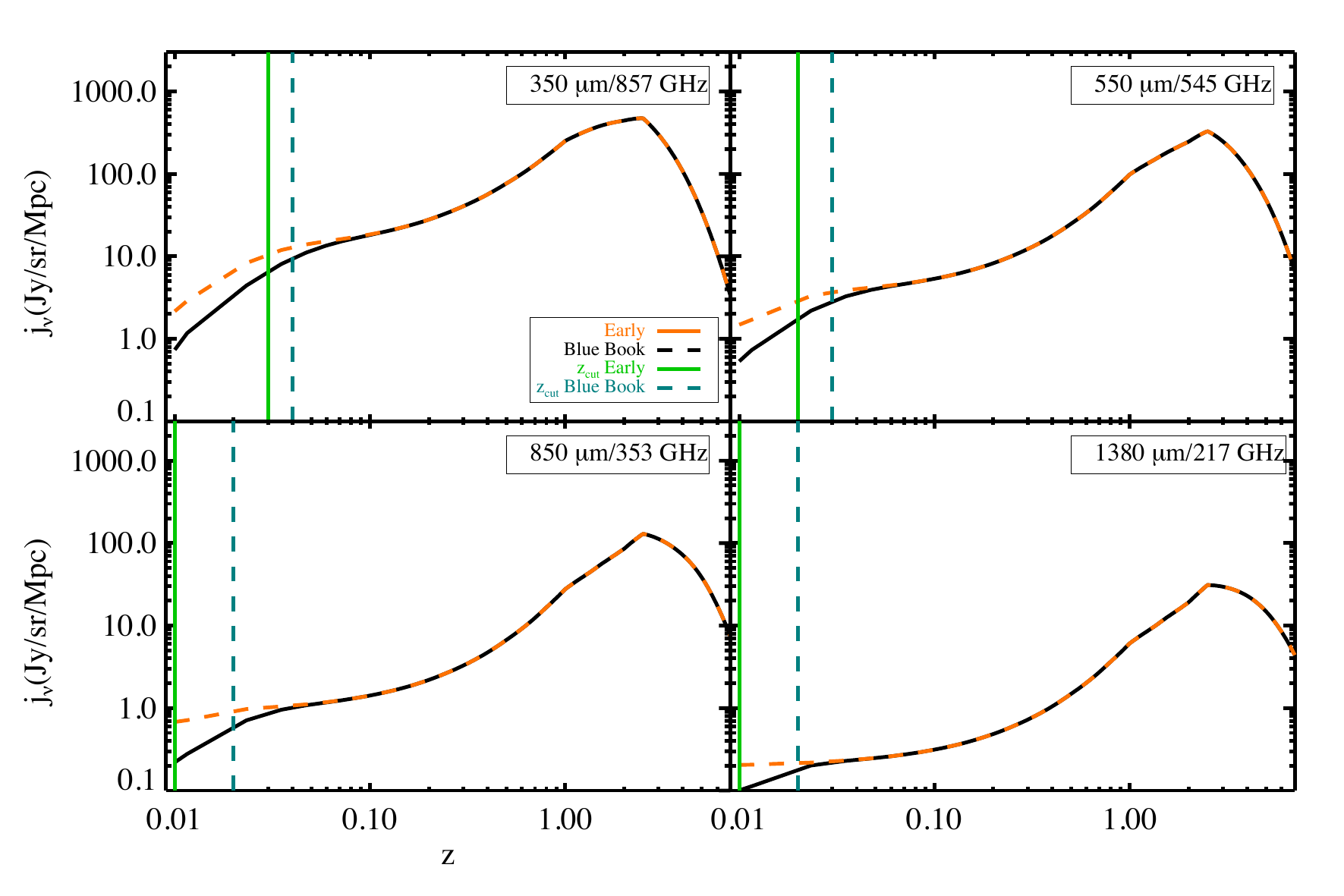} \caption{Emissivities of {\sc Model3} to which flux cuts have been applied. The vertical lines show the redshift cuts. These are illustrated both for ERCSC (orange) and {\tt Planck} Blue Book (black) flux cuts.} \label{fig:compare_jd_fc} \end{figure*}

\begin{figure*}\centering \includegraphics[scale=0.8]{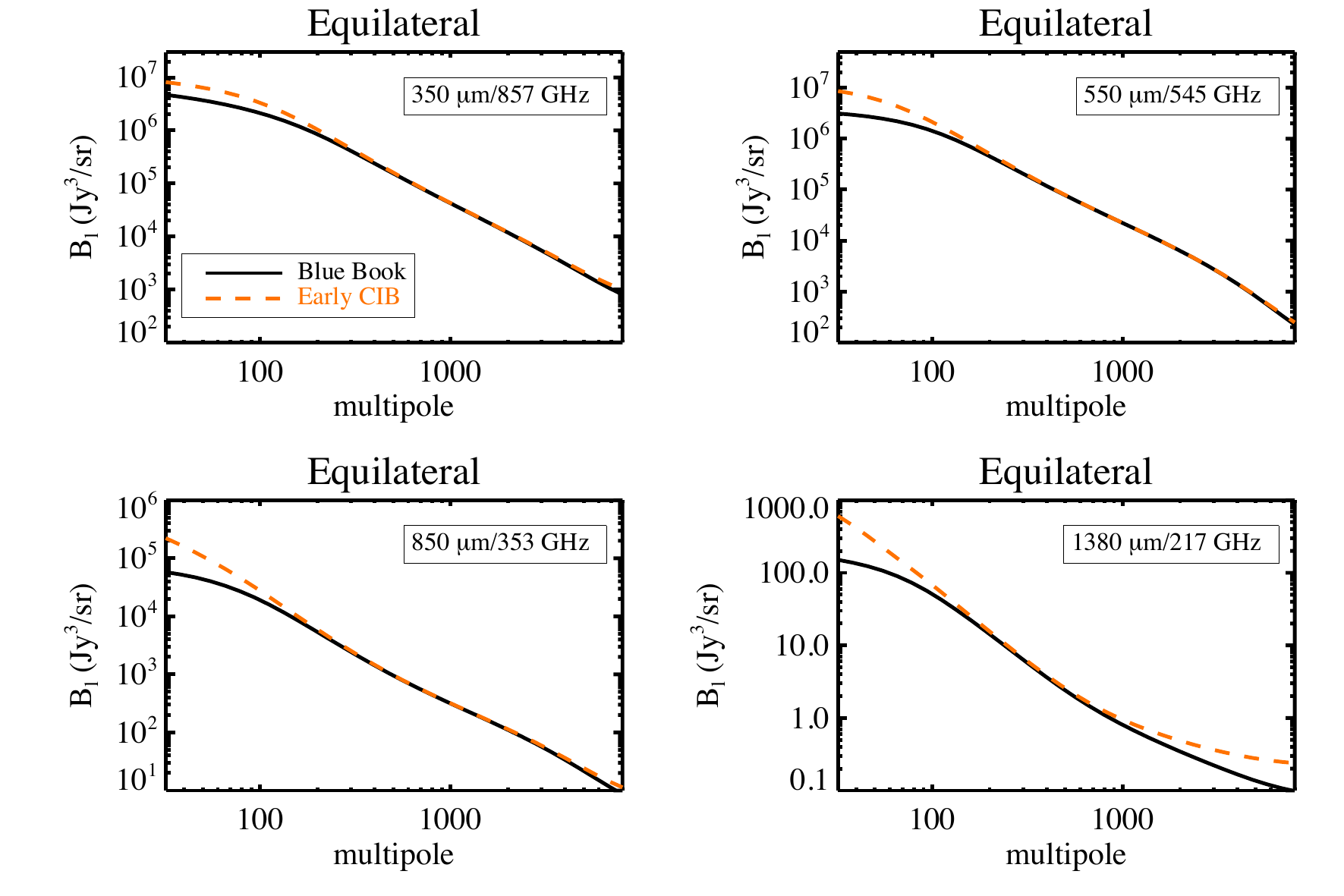} \caption{Equilateral bispectra computed using the two sets of flux cuts. The black and orange lines are the bispectra computed with {\tt Planck} Blue Book and with the ERCSC flux cuts, respectively.} \label{fig:compare_bl_fc} \end{figure*}

When comparing CIB observations and models, one has to account for the so-called flux cut, that is the flux up to which galaxies are resolved as point sources and below which the IR emission is in form of unresolved anisotropies. This limit in flux depends on the resolution of the instrument and its sensitivity, thus on the level of confusion of the data as well as on the contamination of Galactic cirrus. Resolved point sources are most often masked for CIB studies and are mostly at low redshift, while the clustering is dominated by faint higher redshift sources which make the bulk of the anisotropies.  The prediction of any statistics of the CIB, and in particular the power spectrum and the bispectrum, must hence account for the flux cut. It is straightforward to implement the flux cut in the mean emissivities, but one should also account for it in the 3D power spectrum/bispectrum as fewer galaxies are present in the map, particularly at low redshift. This effect must be taken into account in the HOD through the application of a redshift cut associated with the flux cut (see Paper1).

We consider here two sets of flux cuts, a conservative one as given in the ERCSC \citep[][Early Release Compact Source Catalogue of {\tt Planck}]{2011A&A...536A...7P} dedicated to CIB fields where there is not much contamination by Galactic cirrus \citep{2011A&A...536A...7P} and an optimistic one as quoted in {\tt Planck} Blue Book. Flux and associated redshift cuts are listed in Table \ref{tab:flux_cut}. The cuts are applied to both the emissivity and the HOD.

\begin{table*}\centering 
\begin{tabular}{cccccc} \hline\hline 
Wavelength (\um) &Frequency (GHz) & $S^1_\mathrm{cut}$ (mJy) &$z_{\mathrm{cut}}^1$& $S^2_\mathrm{cut}$ (mJy)&$z_{\mathrm{cut}}^2$\\
 \hline 
350 &857 &813 &0.03 & 300 &0.04\\ 
550 &545 &471 &0.02 & 180 &0.03\\ 
850 &353 &249 &0.01 & 75 &0.02\\ 
1380 &217 &280 &0.01 & 37 &0.02\\ 
\hline 
\end{tabular}
 \caption{The flux cuts and associated redshift cuts considered in the present study. $S^1_\mathrm{cut}$ and $z_{\mathrm{cut}}^1$ are given for the ERCSC 10$\sigma$ flux cuts \citep{2011A&A...536A...7P}. $S^2_\mathrm{cut}$ and $z_{\mathrm{cut}}^2$ are the flux cuts at 3$\sigma$ quoted in the {\tt Planck} Blue Book.} 
\label{tab:flux_cut} 
\end{table*}

Making use of {\sc Model3}, we investigate how the bispectrum varies with the two sets of flux/redshift cuts. Fig. \ref{fig:compare_jd_fc} shows the flux-cut emissivities as well as the limits in redshift we consider. The difference between the two sets of emissivities lies at low redshift, as expected. At $z=0.01$, it reaches factors of 2.5 and 1.8 at  350 and 1380 \um, respectively. We can also note that the redshift range on which this difference applies decreases with the wavelength. Indeed, the contribution to the anisotropies of the CIB from low-redshift galaxies decreases with the wavelength \citep{2008A&A...481..885F}. We then compare, for illustration, the equilateral bispectra derived with both sets of flux/redshift cuts as shown in Fig. \ref{fig:compare_bl_fc}. The squeezed bispectrum displays the same behaviour. The main difference between these bispectra lies at large angular scales, up to $\ell\sim250$ at 350 \um~and $\ell\sim120$ at 1380 \um. The largest difference, up to a factor four, is between the two bispectra at 1380 \um. At the longest wavelength, we also note a difference at small angular scales, $\ell \geq 2000$, because of the 1-galaxy shot noise which is higher for the ERCSC flux cuts. Indeed the 1-galaxy shot noise depends strongly on the flux cut as shown in Eq. \ref{eq:shot1g}.\\
We examined each separate term of the bispectrum and found that these differences at large angular scale are mostly due to the 1-halo term which is the most sensitive to the redshift/flux cut, whereas the 2-halo and 2-galaxy shot noise terms are only slightly affected, displaying a variation of $\sim5\%$  at large angular scales. We also investigated the separate effect of respectively the flux cut and the redshift cut on the bispectrum. Although both cuts have a noticeable effect, we found that the differences in Fig.\ref{fig:compare_bl_fc} are mainly due to the flux cuts applied to the emissivities and not to the redshift cuts applied to the HOD.\\
We have seen in Sect, \ref{par:redshift_distrib} that low redshift galaxies dominate the bispectrum at all wavelength. We investigate to which extent we can remove this contribution by applying a more drastic flux cut. Indeed instruments such as {\tt SPT} and {\tt Herschel} have higher resolutions that enable the application of lower flux cuts and thus the removal of more low redshift sources. We apply a flux cut of 50 mJy as used by \citet{2012arXiv1208.5049V} for {\tt Herschel/SPIRE} data. The relative contributions of each redshift bin only vary of a few percents at wavelengths longer than 550 \um, however the situation strongly changes at 350 \um~as shown in Fig. \ref{fig:contrib_z_bl_350_low_fc}. This wavelength, compared to the others, is the most sensitive to low redshift and thus to the flux cut. A lower flux cut reduces the contribution of the low $z$ range. For instance, at $\ell = 93$ for the equilateral configuration, the relative contribution of the slice $0<z<0.7$ drops from 87 \% to 41 \% while that at $0.7<z<1.5$ and $1.5<z<3$ increase from 8 \% to 33\% and from 6 to 24 \%, respectively. The highest redshift bin contributes to less than 1 \% in both case. The application of a low flux cut limits the contribution of low redshift galaxies only at 350 \um. At longer wavelength, the bispectrum keeps being dominated by these galaxies regardless of the flux cut. 

\begin{figure}\centering 
\includegraphics[scale=0.45]{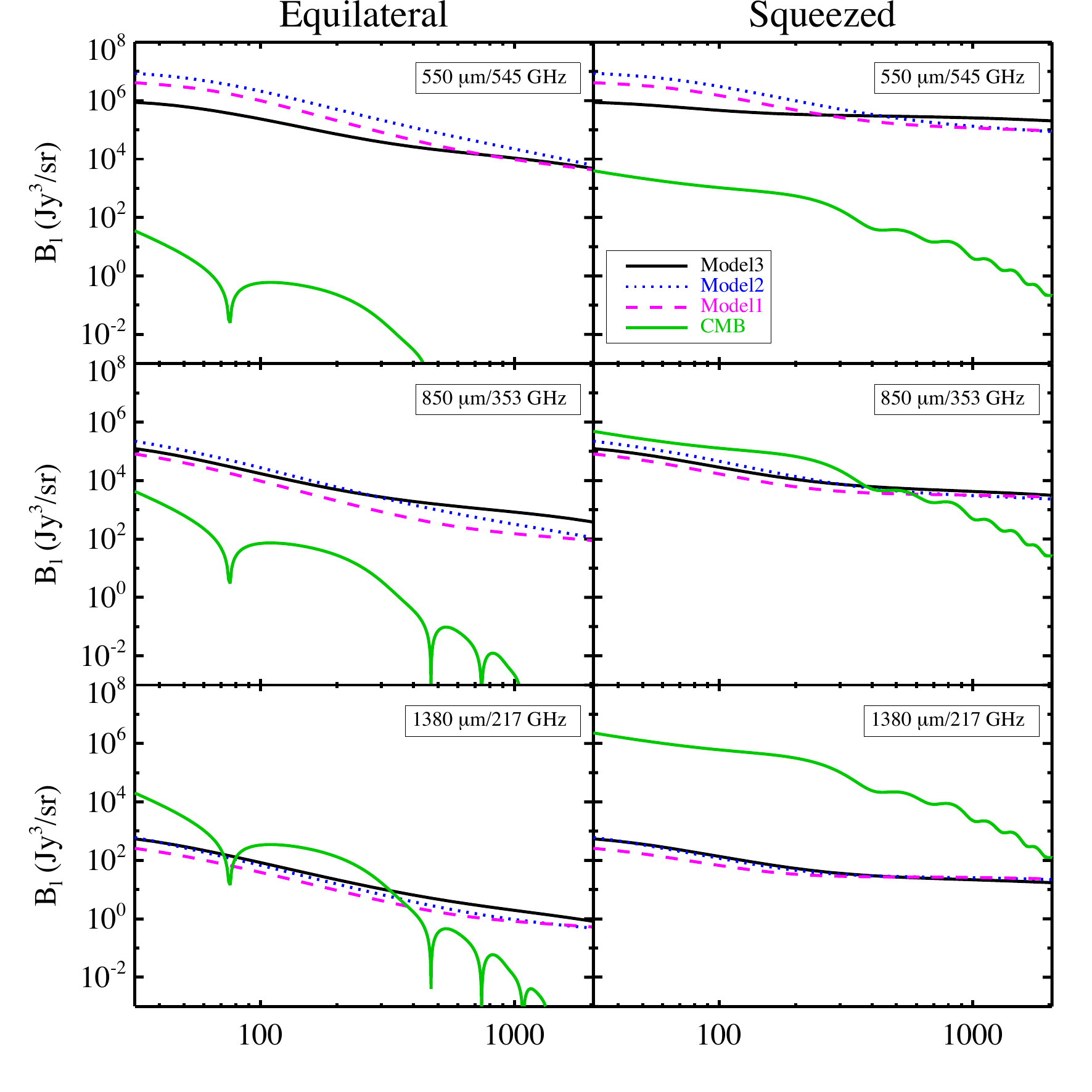} 
\caption{Equilateral and squeezed bispectra of DSFG compared to the CMB bispectrum (green).} 
\label{fig:compare_dsfg_cmb} 
\end{figure}


\section{Contamination of the CMB bispectrum and constraining the HOD parameters}\label{par:delta_fnl}

\begin{figure*}\centering 
\begin{tabular}{cc}
\includegraphics[scale=0.43]{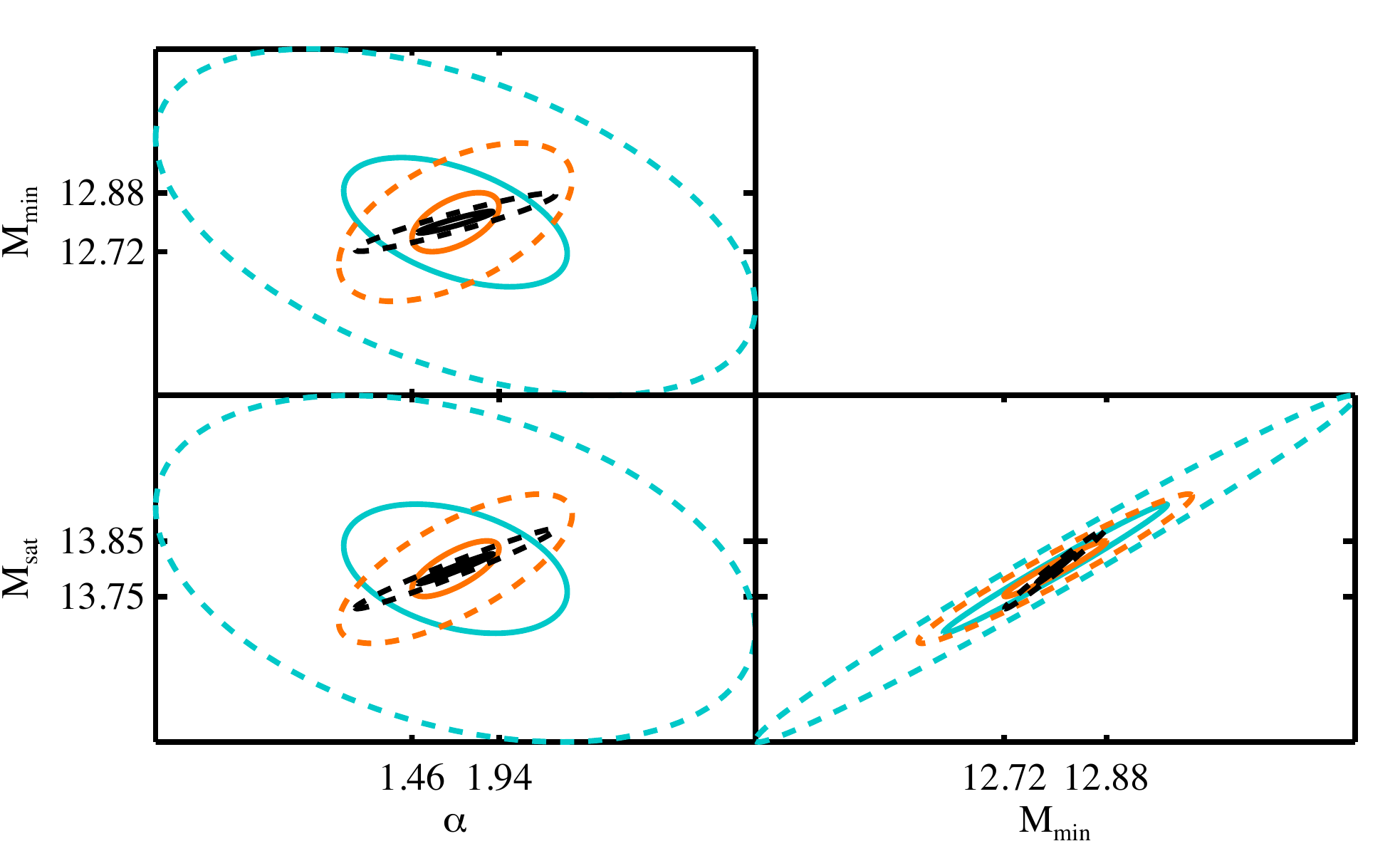}                           &\includegraphics[scale=0.43]{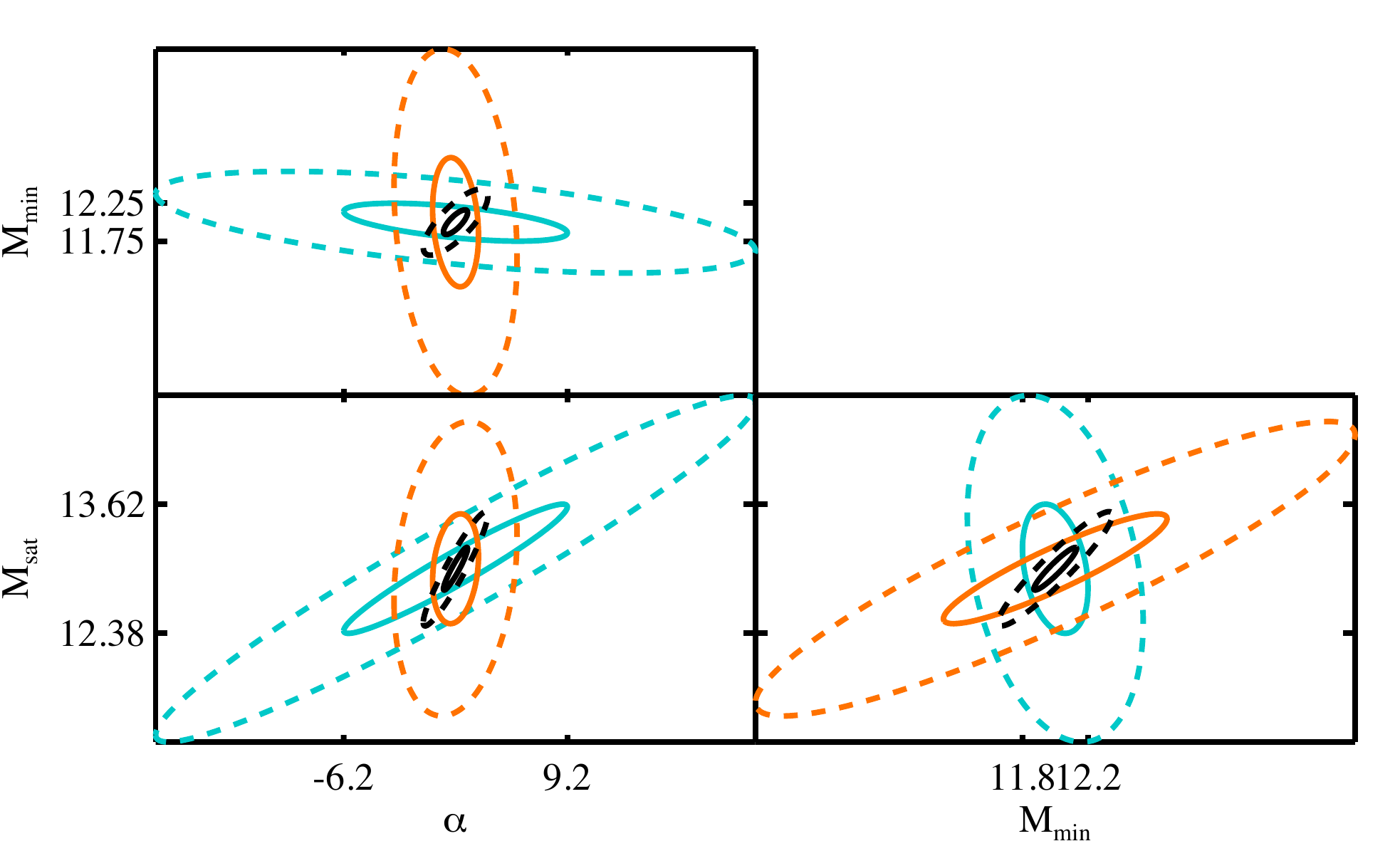} \\
\includegraphics[scale=0.43]{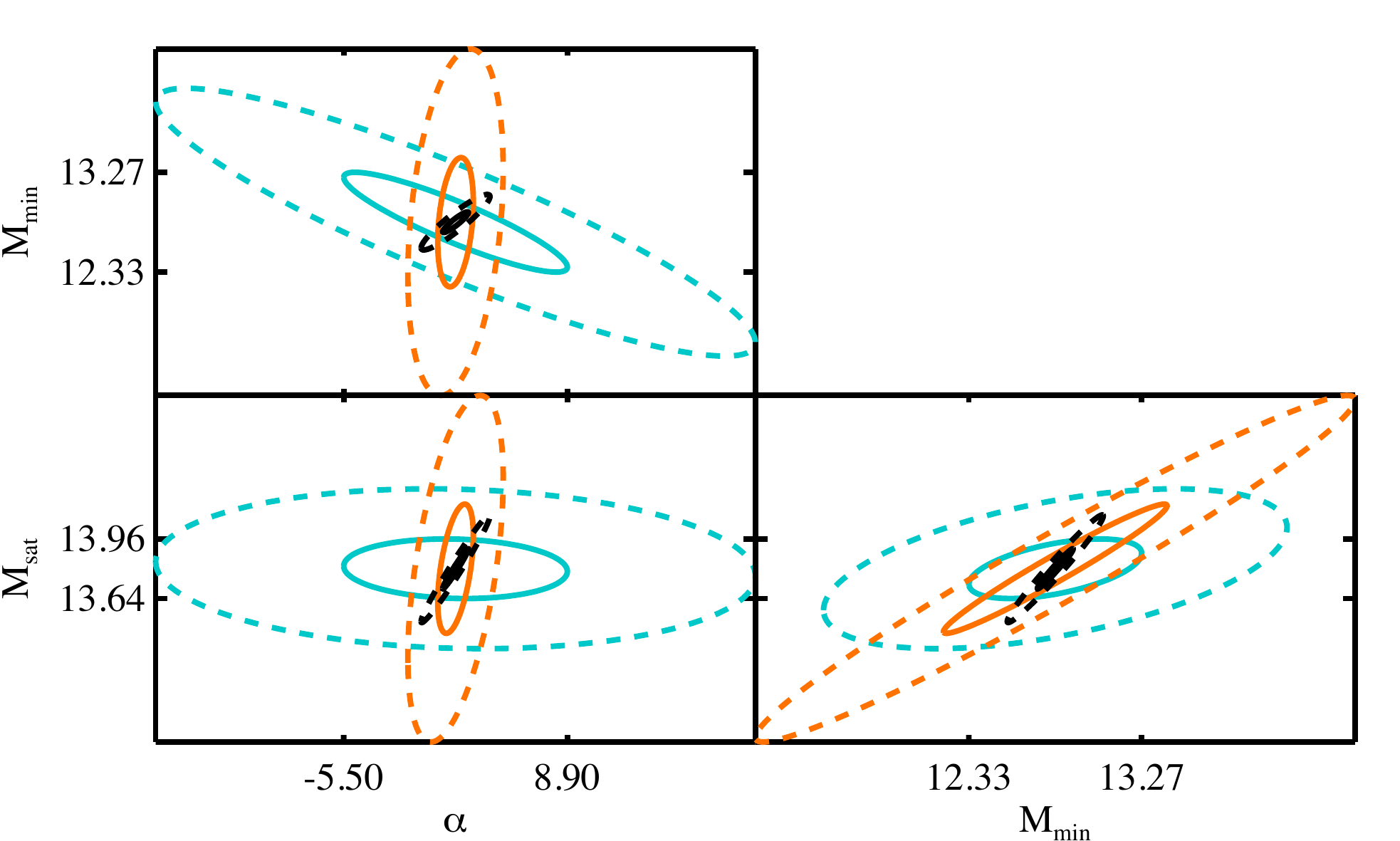} & \includegraphics[scale=0.43]{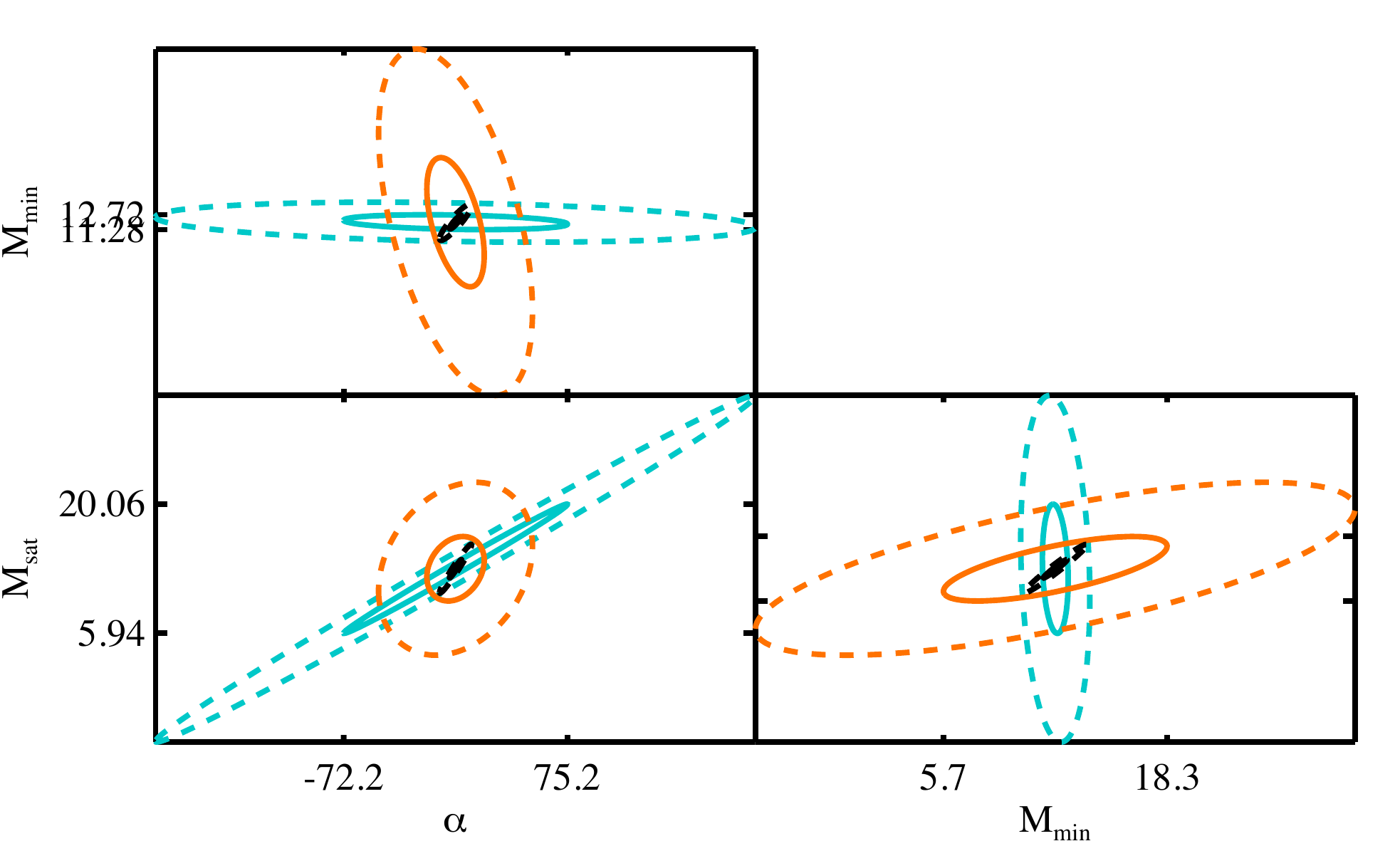} \\
\end{tabular}
\caption{Confidence ellipses for HOD parameters coming from the power spectrum (light blue), the bispectrum (orange) and the combination of both (black) at 350 and 1380 \um~on the left and right panels, respectively. Top panels are for the ideal case with only instrumental noise added to the measurements while lower panels are for the realistic case that includes foregrounds residuals. Solid lines are the confidence ellipses at 1$\sigma$ and the dashed lines are those at 2$\sigma$.} 
\label{fig:ellipse_hod_combined} 
\end{figure*}

\citet{2012MNRAS.421.1982L} studied the contamination of IR galaxies to the CMB bispectrum by computing the CIB anisotropies bispectrum with a prescription grounded on their power spectrum. They first showed that both the CMB and CIB bispectra are maximum in the squeezed configuration. Moreover, they estimated the contamination of the local CMB non-Gaussianity by computing the bias on \fnl~induced by IR galaxies, $\Delta f_\mathrm{NL}$. They showed that the IR galaxies produce a negative bias which becomes important for {\tt Planck}-like resolution and at low wavelengths. For instance, $\Delta f_\mathrm{NL}\sim-6$ at 1380 \um~and $\Delta f_\mathrm{NL}\sim-60-70$ at 850 \um. Most of the signal is due to the clustering of unresolved sources that make the CIB. Therefore they conclude that the bias $\Delta f_\mathrm{NL}$ is not reduced when applying a lower flux cut and, in some cases, it even increases because of the reduction of the shot noise term.\\
Here, we simply compare CIB and CMB bispectra at {\tt Planck} wavelengths. to this end, we assume primordial non-Gaussianity of the local type with $f_\mathrm{NL}=1$. Fig. \ref{fig:compare_dsfg_cmb} shows the comparison of the equilateral and squeezed bispectrum of the CMB to our predictions using the three models of galaxies investigated in the present study. We do not display the comparison at 350 \um~as the CIB bispectrum dominates the CMB one by more than eight orders of magnitude. \\
Concerning the equilateral configuration, the CIB bispectrum highly dominates that of the CMB at 550 \um~as there are 4 orders of magnitude between both. At longer wavelengths, the difference decreases and at 1380 \um~both bispectra have similar amplitudes. The contamination of the CMB bispectrum by that of the CIB thus decreases with the wavelength. The situation changes with respect to the squeezed configuration. At 550 \um, the CIB is still stronger by, at least, two orders of magnitude. At 850 \um, the CMB bispectrum is slightly higher than that of the CIB and at 1380 \um, the CIB bispectrum is dominated by that of the CMB by 3 orders of magnitude. In terms of CIB studies, the equilateral bispectrum is more appropriate to avoid the contamination by the CMB and we confirm that both CMB and CIB bispectra peak in the squeezed configuration .\\
A useful application to the measurement of the non-Gaussianity of CIB anisotropies is to constrain the HOD parameters. \citet{2012A&A...537A.137P} showed, using several wavelengths simultaneously, that even if the HOD parameters  are, on the whole, well constrained, \Mmin~and \Msat~are strongly degenerated. We carry a Fisher analysis to compare, first, how power spectra and bispectra measurements constrain HOD parameters alone and second, how the degeneracies are broken when combining both probes. We cannot use the four wavelengths together as the HOD parameters are different. Therefore we compute Fisher matrices for each wavelength using {\sc model3} and assuming an optimistic fraction of the sky, 50\%, over which spectra and bispectra are available. We bin both the power spectrum and the bispectrum, using 32 bins from $\ell=32$ to $\ell=2048$ with $\Delta\ell=64$. We set $\ell_b$ the mean multipole over a bin. The Fisher matrix of the power spectrum~ is:
\be F_{ij}^{C_{\ell_b}} = \sum_{\ell_b}\frac{1}{\sigma_{\ell_b, C_{\ell_b}}^2}\frac{\dd C_{\ell_b}}{d\theta_i}\frac{\dd C_{\ell_b}}{d\theta_j} f_\mathrm{sky}
\ee
where $\theta_{\mathrm{k}}$ is the k-th parameter and $\sigma_{\ell_b,C_{\ell_b}}$ is the error bar accounting for cosmic variance and instrumental noise: 
\be
\sigma_{\ell,C_{\ell_b}} = (C_{\ell_b} B_{\ell_b}^2+N_{\ell_b})\sqrt{\frac{2}{(2\ell_b+1)\Delta\ell}}
\ee
Here $N_\ell$ is the level of the instrumental noise \citep{2011A&A...536A..18P}, $B_\ell^2$ the power spectrum of the point spread function and $f_\mathrm{sky}$ is the sky fraction.\\ 
The computation of the Fisher matrix is slightly different for the bispectrum. Indeed one needs to compute the full covariance matrix on all the possible configurations of the bispectrum $\mathrm{Cov}(b_{\ell_{123}},b_{\ell'_{123}})$ to compute the associated Fisher matrix. This covariance matrix contains Gaussian and non-Gaussian contributions~:
\be
\mathrm{Cov}(b_{\ell_{123}},b_{\ell'_{123}}) = \mathrm{Cov}_\mathrm{G} + \mathrm{Cov}_\mathrm{NG}
\ee
with the Gaussian contribution being diagonal and leading to an error contribution $\sigma_{\ell_{123}}$~:
\bea
\sigma^2_{\ell_{123}} &=& (C_{\ell_1}B_{\ell_1}^2+N_{\ell_1}) (C_{\ell_2}B_{\ell_2}^2+N_{\ell_2}) (C_{\ell_3}B_{\ell_3}^2+N_{\ell_3})\nonumber\\
&\times&\frac{1}{N_\mathrm{tri}(\ell_1,\ell_2,\ell_3)}
\eea
where $N_\mathrm{tri}(\ell_1,\ell_2,\ell_3)$ is the number of possible configurations of the triplet $(\ell_1,\ell_2,\ell_3)$.\\
The non-Gaussian contribution to the covariance matrix is~:
\bea
\mathrm{Cov}_\mathrm{NG}(b_{\ell_{123}},b_{\ell'_{123}})&=& \frac{1}{\Delta\ell} b_{\ell_{123}}b_{\ell_{123}'}B_{\ell_1}B_{\ell_2}B_{\ell_3}B_{\ell'_1}B_{\ell'_2}B_{\ell'_3}\nonumber\\
&\times&[\frac{\delta_{\ell_1\ell'_1}}{2\ell_1+1}+\frac{\delta_{\ell_1\ell'_2}}{2\ell_1+1}+\frac{\delta_{\ell_1\ell'_3}}{2\ell_1+1}\nonumber\\
&+&\frac{\delta_{\ell_2\ell'_1}}{2\ell_2+1}+\frac{\delta_{\ell_2\ell'_2}}{2\ell_2+1}+\frac{\delta_{\ell_2\ell'_3}}{2\ell_2+1}\nonumber\\
&+&\frac{\delta_{\ell_3\ell'_1}}{2\ell_3+1}+\frac{\delta_{\ell_3\ell'_2}}{2\ell_3+1}+\frac{\delta_{\ell_3\ell'_3}}{2\ell_3+1}]
\eea
The Fisher matrix is then
\be
F_{ij}^{b_\ell} = {}^t \! X_{\theta_i} \mathrm{Cov}(b_{\ell_{123}},b_{\ell_{123}'})^{-1}X_{\theta_j}f_\mathrm{sky}
\ee
where $X_{\theta_k}$ is the vector that contains the derivatives of the bispectrum with respect to the parameter $\theta_k$. \\ 
We consider two cases. First, an ideal case, we assume no foreground residuals in the measurement of the CIB anisotropy power spectra and bispectra. Only instrumental noise is added to the CIB measurements.  Second, we consider a more realistic case in which contributions from CMB and dust residuals are also included to the power spectra and bispectra. \\
The upper panels of Fig. \ref{fig:ellipse_hod_combined} show confidence ellipses at 350 and 1380 \um~derived from the power spectrum (light blue), the bispectrum (orange), and the combination of both (black) for the ideal case. At 350 \um, the constraints induced by the bispectrum are tighter than those from the spectrum. Indeed, the bispectrum is much more sensitive to the variations of the HOD parameters than the power spectrum as shown in Paper1. In addition, the directions of  degeneracy from the power spectrum and bispectrum are orthogonal for the pairs (\alphas,\Mmin) and (\alphas,\Msat) leading to a break of the degeneracies and to a significant improvement of the accuracy when combining both data sets. The situation is different for the pair (\Mmin,\Msat) for which the directions of degeneracy from both the power spectrum and the bispectrum are the same. The improvement in the constraints on this pair of parameters is mostly driven by the higher sensitivity of the bispectrum to mass. At 1380 \um, degeneracies are broken for the three pairs of parameters leading to significantly improved constraints on the HOD parameters. For instance, 1$\sigma$ error bars on \alphas~given by \cl~and \bl~alone are poor, $\sim500$\% and $\sim100$\%, respectively. It is reduced to $\sim50$\% when power spectrum and bispectrum are combined. Note that the large errors bars at long wavelengths are due to instrumental noise. \\
A more realistic case is to consider residuals of foregrounds in the measured power spectra and bispectra. Following \citet{2013arXiv1309.0382P}, we assume 10 \% contamination by the CMB to the CIB power spectrum and we discard from the analysis low multipoles where Galactic dust has a non-negligible contribution to the power spectrum and to the bispectrum. We assume a null non-Gaussianity as \fnl~has been shown to be close to zero \citep{2013arXiv1303.5076P}. The lower panels of Fig. \ref{fig:ellipse_hod_combined} show the confidence ellipses at 350 and 1380 \um~including foreground residuals. Parameters constraints are, of course, poorer for both bispectra and power spectra. Nevertheless the orthogonality of the directions of degeneracy is preserved significantly leading to  improved constraints, still poorer than the ideal case. For instance, at 350 \um~the 1$\sigma$ error bar on \alphas~given by the combination of both data sets is 12\% in the ideal case and increases to 49\% when including foregrounds residuals. The situation is similar at 550 and 850~\um~however, at 1380 \um~the constraints are highly worsened because the CMB is dominant at that wavelength.. The 1$\sigma$ error bar on \alphas~rises from 55 to 220 \% from the ideal to the realistic case.

 The constraints on the HOD parameters heavily depend on the fraction of sky available for the measurement of both power spectrum and bispectrum. This is illustrated by varying the sky fraction from 10\% to a very optimistic case of 70\% of sky. Changing $f_{\mathrm{sky}}$ does not change the degeneracy directions, only the error bars on the parameters, which scale as $\propto\sqrt{f_\mathrm{sky}}$.  These error bars are listed in Table~\ref{tab:eb_fsky} as percentages of the fiducial HOD parameter values for each wavelength and for both ideal and realistic cases. The slope of the number of satellites, \alphas~is the most affected by the addition of foreground residuals, error bars get poorer by a factor $\sim$5 at most. The error bars on \Mmin~and \Msat~are also affected, by a factor $\sim3$ at most, but they remain excellent. In both cases, the small error bars on \Mmin~and \Msat~are striking, at the maximum of order of few percents up to 850 \um. This comes from the orthogonality of the directions of the degeneracies from \cl~and \bl. Constraints at 1380 \um~are highly worsened when adding foregrounds residuals because of the high amplitude of the CMB at this wavelength. We also note that 50\% of the sky is sufficient to constrain accurately the HOD parameters.

\begin{table*}
  \centering 
\begin{tabular}{cccccc|ccc} 
\hline\hline
&&& \multicolumn{3}{c}{Ideal case}&\multicolumn{3}{c}{Realistic case}\\
\hline
 Wavelength &Frequency & $f_\mathrm{sky}$ &\alphas &$\log$\Mmin & $\log$\Msat          &\alphas &$\log$\Mmin & $\log$\Msat \\ 
 \um & GHz &&&&\\  
\hline \multirow{4}{*} {350} &\multirow{4}{*}{857} & 10\% & 27.3\% & 0.5\% & 0.5\%          & 109\%  &1.7\%              &1.8\% \\  
&                                                                            & 30\% & 15.7\% & 0.3\% & 0.3\%          &  63\%   & 1\%                &1\%\\  
&                                                                            & 50\% & 12.2\% & 0.2\% & 0.2\%          &  49\%   &0.8\%              &0.8\% \\  
&                                                                            & 70\% & 10.3\% & 0.2\% & 0.2\%          & 41\%    &0.6\%              &0.7\%\\ 
\hline 
\multirow{4}{*} {550} &\multirow{4}{*}{545} & 10\% & 10.3\% & 0.3\% & 0.2\%                     &50\%     &1.3\%              &1.2\%\\
  &                                                               & 30\% & 6.0\% & 0.2\% & 0.1\%                       &29\%     &0.7\%              &0.7\%\\ 
 &                                                                & 50\% & 4.6\% & 0.1\% & 0.1\%                       &22\%     &0.6\%              &0.5\%\\  
&                                                                 & 70\% & 3.9\% & 0.1\% & 0.1\%                       &19\%     &0.5\%              &0.4\%\\ 
\hline 
\multirow{4}{*} {850} &\multirow{4}{*}{353} & 10\% & 22.7\% & 0.9\% & 0.8\%                     &72\%     &2.4\%              &2.4\%\\  
&                                                                 & 30\% & 13.1\% & 0.5\% & 0.4\%                     &42\%      &1.4\%             &1.4\%\\  
&                                                                 & 50\% & 10.2\% & 0.4\% & 0.3\%                     &32\%      &1\%                &1\%\\  
&                                                                 & 70\% & 8.6\% & 0.3\% & 0.3\%                       &27\%      &0.9\%             &0.9\%\\ 
\hline
 \multirow{4}{*} {1380} &\multirow{4}{*}{217} & 10\% & 123.7\% & 3.0\% & 3.5\%                &582\%    &12\%              &17\%\\ 
 &                                                                   & 30\% & 71.4\% & 1.7\% & 2.0\%                  &337\%    &7\%                & 10\%\\  
&                                                                    & 50\% & 55.3\% & 1.3\% & 1.6\%                  &260\%    &5\%                &8\%\\  
&                                                                    & 70\% & 46.7\% & 1.1\% & 1.3\%                  &220\%     &5\%                &7\%\\ 
\hline 
\end{tabular} 
\caption{1$\sigma$ error bars on the HOD parameters when combining power spectrum and bispectrum data as a function of the available sky fraction at different wavelengths in both the ideal and realistic cases. The fiducial values of the HOD parameters are listed in Table \ref{tab:best_fit}.} 
\label{tab:eb_fsky} 
\end{table*}


\section{Conclusion}\label{par:ccl}

We have presented a model of the bispectrum of unresolved dusty star-forming galaxies in the Cosmic Infrared Background. It is developed in the framework of the halo model and a halo occupation distribution interfaced with a model of evolution of galaxies. We have predicted bispectra of the CIB anisotropies at {\tt Planck} wavelengths for three recent models of evolution of galaxies. Interestingly, these models do not predict the same amount of non-Gaussianity.\\
Surprisingly, regardless of the wavelength, bispectra are dominated by low redshift galaxies ($z<0.7$) whereas it is not the case for the power spectrum. Indeed, for the latter, the contributions of low redshift galaxies decreases as the wavelength gets longer while the contribution of higher redshift galaxies increases. Nevertheless, even if bispectra are dominated by low redshift galaxies, we still recover the trend observed for the power spectrum below that dominant contribution. This strong contribution of the low redshift bin galaxies is due to the 2D projection of the 3D bispectrum.  In order to take into account the flux cut applied to CIB maps for the removal of resolved point sources, we introduced a redshift cut in the HOD. By applying a very low flux cut, for instance one typical of {\tt Herschel} or {\tt SPT}, we are not able to remove a large fraction of the contamination of low redshift galaxies and thus access to the contribution from higher redshift galaxies. \\ 
We have investigated the mass and redshift dependence of each term of the bispectrum and as a function of the wavelength. We do recover similar results to the power spectrum case. The four terms of the bispectrum that depend on the HOD do not have the same mass dependence. For instance, the 1-halo term is dominated by high mass halos up to a redshift of 3 whereas the main contribution to the 2-halo term is from high mass halos at low redshift and the dominating mass shifts to intermediate halo masses at higher redshifts. The contribution of the 3-halo term is mainly due to galaxies that lie in intermediate mass halos in the redshift range $z=[0,4]$.\\
We compare our predictions of the bispectrum of CIB anisotropies to the bispectrum of the CMB in the case of \fnl=1. We do confirm that both bispectra are maximum in the squeezed configuration. It might be complex to disentangle between them at wavelengths longer than 850 \um if the CMB cannot be removed properly. \\
The clustering part of the galaxy model is fully parametric which enables us to carry a Fisher analysis to investigate to which extent the model parameters are constrained with the power spectrum, bispectrum alone and when combining both. We consider one ideal case including only the CIB and the instrumental noise and a more realistic case in which foreground residuals are added. In both cases, when combining both data sets, directions of degeneracy of \cl~and \bl~are usually orthogonal which improves significantly the constraints on the HOD parameters. We show that \cl~and \bl~measurements over a fraction of the sky of 50\% provides between 0.2 \% and 10 \% error bars on the HOD parameters in the ideal case and between 0.5 and 50 \% for the realistic case up to 850 \um.\\


\section*{Acknowledgments}
The authors thank B. B\'ethermin for useful discussions and for providing us the emissivities from \citet{2012ApJ...757L..23B}. The authors also wish to acknowledge  our referee S. Prunet, G. P. Holder, C. Porciani, C. Schimd, E. Sefusatti, G. Lagache, J. Tinker, and A. Wetzel for useful discussions and M. Langer for a thorough reading of the manuscript.


\bibliographystyle{mn2e}

\bibliography{Biblio} \label{lastpage}

\end{document}